\documentclass[amssymb,prb,twocolumn,showpacs]{revtex4-1}
\usepackage{epsfig}
\usepackage{dcolumn}
\usepackage{amsmath,amssymb}
\let \Re \relax
\let \Im \relax

\def \beq {\begin{equation}}
\def \edq {\end{equation}}
\def \bes {\begin{subequations}}
\def \eds {\end{subequations}}
\def \beqn {\begin{equation*}}
\def \edqn {\end{equation*}}

\def \dag {\dagger}

\def \up {\uparrow}
\def \down {\downarrow}
\def \eps {\epsilon}
\def \sm {\sigma}
\def \bsm {\bar{\sigma}}

\def \veps {\varepsilon}

\def \calh {{\cal{H}}}

\def \calg {{\cal{G}}}

\def \calo {{\cal{O}}}

\def \calt {{\cal{T}}}
\def \calr {{\cal{R}}}

\def \hatO {\hat{O}}

\def \wteps {\widetilde{\veps}}

\def \wG {\widetilde{\Gamma}}

\providecommand{\nbraket}[1]{\langle#1\rangle}
\providecommand{\dbraket}[1]{\langle\langle#1\rangle\rangle}

\DeclareMathOperator{\Re}{Re}
\DeclareMathOperator{\Im}{Im}

\begin{document}

\title{Magnetoasymmetric transport in a mesoscopic interferometer:
From the weak to the strong coupling regime}
\author{Jong Soo Lim}
\affiliation{Departament de F\'{\i}sica,
Universitat de les Illes Balears, E-07122 Palma de Mallorca, Spain}
\author{David S\'anchez}
\affiliation{Departament de F\'{\i}sica,
Universitat de les Illes Balears, E-07122 Palma de Mallorca, Spain}
\author{Rosa L\'opez}
\affiliation{Departament de F\'{\i}sica,
Universitat de les Illes Balears, E-07122 Palma de Mallorca, Spain}
\date{\today}

\begin{abstract}
The microreversibility principle implies that the conductance of a two-terminal
Aharonov-Bohm interferometer is an even function of the applied magnetic flux.
Away from linear response, however, this symmetry is not fulfilled and
the conductance phase of the interferometer when a quantum dot is inserted
in one of its arms can be a continuous function of the bias voltage.
Such magnetoasymmetries have been investigated in related mesoscopic systems
and arise as a consequence of the asymetric response of the internal
potential of the conductor out of equilibrium. Here we discuss magnetoasymmetries
in quantum-dot Aharonov-Bohm interferometers when strong electron-electron
interactions are taken into account beyond the mean-field approach.
We find that at very low temperatures
the asymmetric element of the differential conductance
shows an abrupt change for voltages around the Fermi level.
At higher temperatures we recover a smooth variation of the magnetoasymmetry
as a function of the bias. We illustrate our results with the aid of the electron
occupation at the dot, demonstrating that its nonequilibrium component
is an asymmetric function of the flux even to lowest order in voltage.
We also calculate the magnetoasymmetry of the current--current
correlations (the noise) and find that it is given, to a good extent,
by the magnetoasymmetry of the weakly nonlinear conductance term.
Therefore, both magnetoasymmetries (noise and conductance)
are related to each other via a higher-order fluctuation-dissipation
relation. This result appears to be true even in the low temperature
regime, where Kondo physics and many-body effects
dominate the transport properties.
\end{abstract}

\pacs{73.23.-b, 73.50.Fq, 73.63.Kv}
\maketitle

\section{Introduction}\label{sec_intro}
Transport in electric conductors is governed by
fundamental principles when the fields applied
to the system are small. For instance, in the linear
regime microscopic reversibility leads to symmetric
response coefficients, as demonstrated by Onsager.\cite{ons}
In the case of a conductor coupled to two terminals, the linear
conductance $G_0$ is an even function of the magnetic field $B$.\cite{cas}
When the conductor is reduced to typical sizes
less than the phase-breaking length, electron mesoscopic transport
depends on the particular arrangement of the attached probes
in a multiterminal configuration
in such a way that current and voltage terminals must be exchanged
to recover the Onsager symmetry.\cite{but86}
On the other hand, in the mesoscopic regime
interference effects associated to the wave
nature of carriers can be detected in a transport measurement.
A prominent example is an Aharonov-Bohm interferometer
with a quantum dot inserted in one of its arms
for which the linear conductance $G_0$ is periodically
modulated by the externally applied flux.
However, the conductance phase $\delta$,
which can be related to the transmission phase
through the quantum dot,
shows abrupt jumps as a function of the gate voltage\cite{yac95}
since the Onsager symmetry establishes that
$\delta$ can be $0$ or $\pi$ only.\cite{lev95}
Further theoretical\cite{hac96,bru96} and experimental works\cite{yac96,shu97}
have addressed the effect of electron-electron
interactions inside the quantum dot.

Away from linear response, the principle of microscopic reversibility
is, generally, not satisfied and, as a consequence, the two-terminal current,
which consists of linear as well as nonlinear coefficients,
is not a symmetric function of $B$. Recently, this {\em magnetoasymmetry} effect
has been theoretically demonstrated
\cite{san04,spi04,but05,san05,pol06,mart06,and06,pol07,san08,kal09,sza09,her09}
and experimentally verified \cite{rik05,wei06,mar06,let06,zum06,ang07,har08,che09,bra09}.
Magnetoasymmetries arise because the charge response of the system
is, generally, not symmetric
when the field orientation is inverted.\cite{san04,spi04}
Out of equilibrium, the piled-up charge injected from the external
reservoirs is partly balanced by the screening potential of the conductor.
This internal potential is not an even function of $B$
(the Hall potential is a paradigmatic example)\cite{san04},
leading to magnetoasymmetries
seen already in the secod-order coefficients within an expansion of
currents in powers of voltages.\cite{san04}
Now, computation of the internal potential due to long range Coulomb
interaction is a difficult task and requires self-consistency. This calculation
can be achieved within a mean-field scheme, as previous
works have done.\cite{san04,spi04,but05,san05,pol06}
However, the importance of strong
electron-electron correlations
such as those giving rise to the Kondo effect\cite{theoretical,experimental}
has not been clarified yet.
This is the goal we want to accomplish in this work.

Our calculations are also relevant in view of recent developments
that relate the magnetoasymmetries of the current and that of the noise
to leading order in a voltage expansion.\cite{foe08,sai08,san09,uts09}
It has been shown that novel fluctuations relations hold
in the weakly nonlinear regime between
the asymmetric second-order conductance and
the first-order noise susceptibility in terms of a higher-order
fluctuation-dissipation theorem. 
These works explicitly check this relation for specific systems
by treating interactions in the mean-field limit.\cite{foe08,sai08,san09,uts09}
Thus, it is highly desirable to find systems in which the nonequilibrium
fluctuation relations can be checked beyond the mean-field case.

We here consider a quantum dot embedded in one of the arms
of a two-terminal mesoscopic interferometer.
We employ the nonequilibrium Keldysh Green  function formalism
to describe the transport properties of the system and
use an equation-of-motion
technique to calculate both the current and the noise
in the nonlinear regime
including electron-electron correlation terms not present
in the Hartree-Fock (mean-field) approximation.
\begin{figure}
\centering
\includegraphics[width=0.45\textwidth]{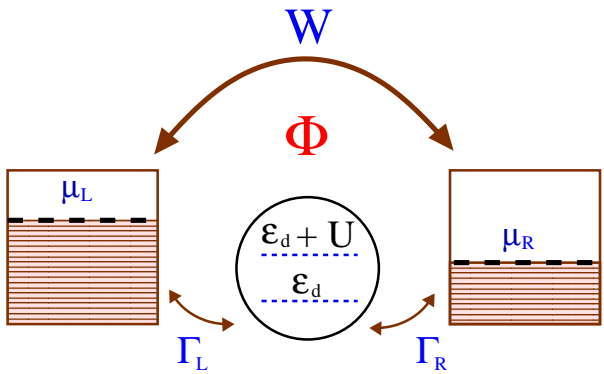} 
\caption{(Color online) Sketch of a mesoscopic interferometer
with a quantum dot inserted in the lower arm.}
\label{fig:interferometer}
\end{figure}

The oscillations in the conductance of an Aharonov-Bohm ring
occur in the mesoscopic regime, where the electrons' phase coherence
is preserved. Interference takes place between electron waves 
that pick up different phases while traversing the two arms of the interferometer
even if the electron does not directly experience the flux enclosed
by the ring. When a finite bias is applied, the phase of the differential
conductance shows a continuous variation with $B$ since the Onsager
symmetry need not hold away from linear response.\cite{bru96,kon02,wie03}
Such phase rigidity
breakings have been recently related to the onset of inelastic
cotunneling of a Coulomb-blockaded quantum dot placed
in one of the arms.\cite{sig07,pul09} The effect is rather generic
as the phase symmetry can be also broken using
microwave fields\cite{sun99} or coupling to phonons.\cite{rom07}

We now give a simple argument
that sheds light on the appearance of magnetoasymmetries
in a quantum-dot Aharonov-Bohm ring.
A sketch of the system is depicted in Fig. \ref{fig:interferometer}.
We model the interaction in the dot
with an on-site charging energy $U$. We assume for simplicity that the dot
contains a single energy level $\varepsilon_d$ which acquires a finite
lifetime $2\hbar/\Gamma$ due to coupling to the external reservoirs.
Then, within the Hartree approximation for the Anderson Hamiltonian\cite{anderson} the
(retarded) Green function for dot electrons with spin $\sigma$ reads,
\begin{equation}\label{eq_retarded_green}
G_{d\sigma}^r(\omega)=\frac{1}{\omega - \veps_d - U\nbraket{n_{d\bsm}}
+ \frac{1}{4}\sqrt{\calt_b} \Gamma \cos\left(\frac{e\Phi}{\hbar}\right) + \frac{i}{2} \widetilde{\Gamma}}\,,
\end{equation}
where $\widetilde{\Gamma}$ renormalizes $\Gamma$ due to the presence
of the nonresonant channel (the upper arm in Fig. \ref{fig:interferometer}).
Its (energy-independent) transmission is denoted with ${\calt_b}$.
Importantly, the renormalization of the level $\veps_d$ is given by two terms.
The term that explicitly depends on ${\calt_b}$ is an {\em even} function
of the magnetic flux $\Phi$ and it fulfills the Onsager symmetry.
However, the term $U\nbraket{n_{d\bsm}}$, proportional to the interaction
strength, is generally not symmetric under field reversal since
the dot occupation $\nbraket{n_{d\bsm}}$ out of equilibrium need not fulfill
the Onsager symmetry. 

Let us expand the dependence of $\nbraket{n_{d\bsm}}$ on the external bias $V$
in powers of $V$,
\begin{equation}\label{eq_occupation}
\nbraket{n_{d\bsm}}=\nbraket{n_{d\bsm}}^{(0)}+\nbraket{n_{d\bsm}}^{(1)}V+\calo(V)^2\,.
\end{equation}
$\nbraket{n_{d\bsm}}^{(0)}$ is the equilibrium charge and must be $B$-symmetric.
The nonequilibrium response of the dot to leading order in $V$ is given by
$\nbraket{n_{d\bsm}}^{(1)}=\partial \nbraket{n_{d\bsm}}/\partial V|_{V=0}$. 
$\nbraket{n_{d\bsm}}^{(1)}$ is then determined from the change in the dot
occupation when a small shift is applied to the leads' electrochemical potential.
Hence, $\nbraket{n_{d\bsm}}^{(1)}$ is a charge susceptibility that
includes information about the screening properties
of the dot and, as such, must be computed self-consistently in the presence of $V$
(e.g., from charge-neutrality condition).\cite{christen}
We below show that processes of charge filling of the dot from the left lead
contribute to the occupation
with a term proportional to $1+\sqrt{\calt_b}\sin (e\Phi/\hbar)$
whereas the contribution from the right lead is proportional
to $1-\sqrt{\calt_b}\sin (e\Phi/\hbar)$. Clearly, the sine terms are not even under $B$-reversal. 
As a result, electron transfer from left to right at a given orientation
of $B$ does not occur with the same probability that the reverse transfer
when the $B$ direction is inverted. In other words,
the injectivity from the left, which is the partial density of states associated to carriers injected from
the left contact,\cite{but93} does not equal
the right injectivity under $B$ reversal.\cite{san04}
As a consequence, one finds  $\nbraket{n_{d\bsm}}^{(1)}\propto \sqrt{\calt_b}\sin (e\Phi/\hbar)$.
Inserting Eq.\ (\ref{eq_occupation}) in Eq. (\ref{eq_retarded_green}), there arises
in the denominator a term proportional to $U V \sqrt{\calt_b}\sin (e\Phi/h)$, which is
responsible, to leading order, for the magnetoasymmetry of the nonlinear conductance.
Note that this term vanishes in the absence of interactions ($U=0$) or at equilibrium
($V=0$) in which cases the Onsager symmetry is recovered. This further demonstrates
that magnetoasymmetries arise as a consequence of the presence of {\em both} interactions
and external bias.

The asymmetric behavior discussed here can be verified experimentally.
Recent experiments with rings show microreversibility violations in the nonlinear regime.
An unexpected even-odd behavior has been revealed by Leturcq {\em et al.}\cite{let06}
In an expansion of the observed current--voltage characteristics,
they find that the odd (even) coefficients are symmetric (asymmetric) under reversal of
$B$. This is surprising since one would expect all coefficients beyond the linear
response to be asymmetric, not only the even ones. Therefore,
it is instructive to derive the conductance series expansion.
The magnetoasymmetric effect appears in the nonlinear regime only
because the internal potential is an asymmetric function of $B$
at finite bias. And this dependence on the internal potential is shown
in the even coefficients, not in the odd ones. A mean-field description gives
this behavior\cite{let06} since the dot potential depends linearly with $V$,
as can be also inferred from our discussion above. Below, we find that the effect
persists even if higher electronic correlations are taken into account.

Microreversibility at linear response also leads to the fluctuation-dissipation
theorem, which relates the dissipative part of the electric transport (the linear conductance)
to the fluctuations at equilibrium (the thermal noise). Since it is clear that microreversibility
is broken beyond the linear response regime, it thus natural to ask whether higher-order
fluctuation relations exist in the presence of an external field.
We find an approximate verification of such relations to next order in the voltage expansion.
In other words, the asymmetric part of the second-order conductance and that of the
linear-order noise are related to each order via a nontrivial fluctuation relation.
Remarkably, our results are interesting because we treat interactions
beyond the Hartree-Fock approximation.

The paper is organized as follows. In Sec.~\ref{sec_sys} we introduce the system
and its Hamiltonian. Section.~\ref{sec_model}
is devoted to the transport properties. We also analyze two approximations: the noninteracting
case and the Hartree approach. Both limits have serious drawbacks and can even lead
to unphysical predictions but we nevertheless include discuss them for pedagogical reasons.
A detailed study of the transport coefficients in the Coulomb blocakde regime
is contained in Sec.~\ref{sec_cb}. We examine the even-odd properties 
of the conductance terms both for zero and nonzero temperatures.
In Sec.~\ref{sec_kondo} we consider the Kondo regime, which is the relevant
scenario at very low temperatures.
We briefly discuss the limits of temperatures lower and higher than
the Kondo temperature and then present
numerical results for the magnetoasymmetric differential conductance
as a function of the background transmission and bias voltage.
In Sec.~\ref{sec_noise} we calculate the noise and show that it is an asymmetric
function of the magnetic flux. Finally, our results are summarized
in Sec. ~\ref{sec_concl}.

\section{Theoretical model}\label{sec_sys}

The mesoscopic interferometer consists of an Aharonov-Bohm ring
with a quantum dot inserted in one of
its arms. The interferometer is coupled to left ($L$) and right ($R$) leads
with continuous energy spectrum $\veps_{k\sigma}$ where $k$ is the wavevector.
An electron can travel either through the nonresonant arm with
probability amplitude $W$ or via the quantum dot with hopping terms $V_{\alpha}$
where $\alpha=L,R$ is the contact index. The spin-degenerate level dot is denoted with $\veps_d$
and the charging energy is $U$. Finally, the magnetic flux $\Phi$ piercing
the ring results in an Aharonov-Bohm phase $\varphi= e\Phi/\hbar$. Hence,
the Hamiltonian reads,
\beq
\calh = \calh_{C} + \calh_{D} + \calh_{T}\,.
\label{eq:model}
\edq
Here,
\begin{multline}
\calh_{C} = \sum_{\alpha=L/R,k,\sigma} \veps_{k\sigma} c_{\alpha k\sigma}^{\dag}c_{\alpha k\sigma} \\
+ \sum_{k,k',\sigma} \left(W e^{i\varphi} c_{Rk'\sigma}^{\dag}c_{Lk\sigma}+{\rm h.c.}\right)\,,
\end{multline}
describes the two leads and the direct channel that couples them.
The gauge is chosen in such a way that an electron wave
picks up the phase $\varphi$ whenever it passes along the upper arm.
The dot electrons obey,
\begin{equation}
\calh_{D} = \sum_{\sigma} \veps_d d_{\sigma}^{\dag} d_{\sigma} + U n_{d\up}n_{d\down} \,,
\end{equation}
while the tunneling Hamiltonian between the reservoirs and the dot is given by,
\begin{equation}
\calh_{T} = \sum_{\alpha=L/R,k,\sigma} \left( V_{\alpha} c_{\alpha k\sigma}^{\dag} d_{\sigma} + h.c.\right)\,,
\end{equation}
where the tunneling amplitudes $V_{\alpha}$ are asummed to be independent of $k$ for simplicity.
The same assumption is made for the direct transmission $W$.

\section{Transport properties}\label{sec_model}
In the stationary limit, the current $I$ can be calculated from
the time evolution of the occupation number of the right contact ($n_R$),
\beq
I = I_R = -e\frac{d\langle n_R\rangle}{dt} = -\frac{ie}{\hbar} \langle [\calh,n_R] \rangle\,,
\edq
where 
$n_R = \sum_{k\sigma} c_{Rk\sigma}^{\dag}c_{Rk\sigma}$.
Using the Keldysh formalism,\cite{mei92} the current becomes
\begin{multline}
I = 
\frac{e}{h} \sum_{p,q,\sigma} \int_{-\infty}^{\infty} d\omega~ \left[
\left( V_R \calg_{d\sigma,Rq\sigma}^{<}(\eps) - V_R^{\ast}\calg_{Rq\sigma,d\sigma}^<(\omega) \right)
\right.
\\
+
\left.
\left( We^{i\varphi} \calg_{Lp\sigma,Rq\sigma}^{<}(\omega) 
- W^*e^{-i\varphi} \calg_{Rq\sigma,Lp\sigma}^<(\omega) \right) \right],
\end{multline}
with the following definitions for the lesser Green fucntion ($G^<$):
\bes
\begin{align}
\calg_{Lp\sigma,Rq\sigma}^{<} &= i\langle c_{Rq\sigma}^{\dag}c_{Lp\sigma} \rangle \,,\\
\calg_{Rq\sigma,Lp\sigma}^{<} &= i\langle c_{Lp\sigma}^{\dag}c_{Rq\sigma} \rangle \,,\\
\calg_{d\sigma,Rq\sigma}^{<} &= i\langle c_{Rq\sigma}^{\dag}d_{\sigma} \rangle \,,\\
\calg_{Rq\sigma,d\sigma}^{<} &= i\langle d_{\sigma}^{\dag}c_{Rq\sigma} \rangle\,.
\end{align}
\eds

In the case of energy-independent couplings or for proportionate couplings,
the expresssion for the current is more conveniently recast in terms
of a generalized transmission function $\calt_{\sigma}(\omega)$
for an electron with spin $\sigma$
and energy $\omega$,
\beq
I = -\frac{e}{h} \sum_{\sigma} \int d\omega~ \calt_{\sigma}(\omega) [f_L(\omega) - f_R(\omega)]\,,
\label{eq:current}
\edq
where $f_L$ and $f_R$ are
the Fermi-Dirac distribution functions in the leads $L$
and $R$, respectively. The transmission reads,\cite{bul01,hof01}
\begin{multline}\label{eq:calt}
\calt_{\sigma}(\omega) = \calt_b + \sqrt{\alpha\calt_b \calr_b} \cos(\varphi) \widetilde{\Gamma} \Re\left[\calg_{d\sigma,d\sigma}^r(\omega)\right] \\
- \frac{1}{2} \left\{\alpha\left[1-\calt_b \cos^2(\varphi)\right] - \calt_b\right\} \widetilde{\Gamma} \Im \left[\calg_{d\sigma,d\sigma}^r(\omega)\right]\,.
\end{multline}
Here, $\calt_b$ is the transmission probability between the two leads
along the direct channel,
\beq
\calt_b = \frac{4\xi}{(1+\xi)^2}\,,
\edq
where $\xi = \pi^2 W^2 \rho_L\rho_R$ with
$\rho_{L(R)}$ the density of states for lead $L(R)$.
The reflection probability $\calr_b$ is determined from $\calr_b = 1 - \calt_b$.
The broadening of the dot level due to hybridization with states of lead $L(R)$ reads
$\Gamma_{L(R)} = 2\pi |V_{L(R)}|^2 \rho_{L(R)}$,
where $\rho_{L(R)}$ is the density of states for lead $L(R)$.
The total linewidth is $\Gamma = \Gamma_L + \Gamma_R$.
In the wide band limit, we take $\rho_{L(R)} = \rho_0$ (and, consequently, $\xi$ and $\Gamma$)
to be energy independent. In the presence of the upper bridge,
the broadening becomes renormalized,
\beq
\widetilde{\Gamma} = \frac{\Gamma}{1+\xi}\,, 
\edq
Finally, the factor $\alpha = 4\Gamma_L\Gamma_R/\Gamma^2$
in Eq.~(\ref{eq:calt}) quantifies the tunneling asymmetry ($0\le\alpha\le 1$).
It yields $\alpha=1$ for symmetric couplings ($\Gamma_L=\Gamma_R=\Gamma/2$).

The problem has thus been reduced to the calculation of
the dot retarded Green function $\calg_{d\sigma,d\sigma}^r(\omega)$.
For later convenience, we introduce the following definitions:
\bes
\begin{align}
\dbraket{A(t),B(t')}^< &= +i\nbraket{B(t')A(t)}\,, \\
\dbraket{A(t),B(t')}^> &= -i\nbraket{A(t)B(t')}\,, \\
\dbraket{A(t),B(t')}^r &= -i\Theta(t-t') \nbraket{[A(t),B(t')]_+}\,, \\
\dbraket{A(t),B(t')}^a &= +i\Theta(t'-t) \nbraket{[A(t),B(t')]_+}\,,
\end{align}
\eds
where $r$ ($a$) stands for ''retarded" ("advanced"). 
Using the operators $A=d_\sigma$ and $B=d_\sigma^\dagger$ we obtain
the dot Green function,
\beq
\calg_{d\sigma,d\sigma}^r(\omega)=\int_{-\infty}^\infty dt\, e^{i \omega t} \dbraket{d(t),d^\dagger(0)}^r\,,
\edq
where we have set $t'=0$ since the Hamiltonian is time independent.
The equation of motion for $\calg_{d\sigma,d\sigma}^r$ is found as,
\begin{multline}\label{eq:eom}
\left[ \omega - \veps_d + \frac{1}{4}\sqrt{\alpha\calt_b} \Gamma \cos(\varphi) + \frac{i}{2} \widetilde{\Gamma}\right] \langle\langle d_{\sm},d_{\sm}^{\dag} \rangle\rangle ^r
= 1 \\+ U \langle\langle d_{\sm} n_{d\bsm},d_{\sm}^{\dag} \rangle\rangle ^r\,.
\end{multline}

\subsection{Noninteracting case}
In the absence of the interaction, one sets $U = 0$ in Eq.~(\ref{eq:eom})
and the retarded Green function is readily computed,
\beq\label{eq:nongreenddr}
\calg_{d\sigma,d\sigma}^{r(0)}(\omega) = \frac{1}{\omega - \veps_d + \frac{1}{4}\sqrt{\alpha\calt_b} \Gamma \cos(\varphi) + \frac{i}{2} \widetilde{\Gamma}}\,.
\edq
Substituting in Eq.~(\ref{eq:calt}), the transmission probability becomes,
\beq\label{eq_caltfano}
\calt_{\sigma}(\omega) =
\frac{\calt_b}{\widetilde{\omega}^2 + \frac{1}{4}\widetilde{\Gamma}^2} \left| \widetilde{\omega} + \frac{q_F \widetilde{\Gamma}}{2} \right|^2\,,
\edq
where
\bes
\begin{align}
\widetilde{\omega} &= \omega - \veps_d + \frac{1}{4}\sqrt{\alpha\calt_b} \Gamma \cos(\varphi)\,, \\
q_F &= \sqrt{\frac{\alpha}{\calt_b}} \left[ \sqrt{\calr_b}\cos(\varphi) + i \sin(\varphi) \right]\,.
\end{align}
\eds
Equation~(\ref{eq_caltfano}) is evidently of the Fano type.\cite{fano}
The Fano antiresonances arise as a consequence of interference between
a direct path channel (the upper arm in Fig. \ref{fig:interferometer}) and a hopping path
via a quasi-localized state (the quantum dot in the lower arm).
As a result, a characteristic asymmetric transmission lineshape
is obtained, which is described with the (generally complex)
Fano parameter $q_F$.
When $\varphi$ is a multiple of $\pi$, the transmission vanishes
at the special energy point given by $\widetilde{\veps}(\veps)=-q_F \widetilde{\Gamma}/{2}$.
On the other hand, for vanishingly small transmission along the direct channel
($\calt_b\to 0$), Equation~(\ref{eq_caltfano}) reduces to the Lorentzian form
of the transmission resonance through a noninteracting dot.

We take a bias $V$ symmetrically applied to the electrodes
($\mu_L = -\mu_R = eV/2$) and insert Eq.~(\ref{eq_caltfano})
in Eq.~(\ref{eq:current}). Next, we expand the current--voltage characteristics,
$I = G_0 V + G_1 V^2 + G_2 V^3 + \cdots=\sum_{n\ge 0} G_n V^{n+1}$.
We find that the even coefficients $G_{2n}$ ($n=0,1,2\ldots$)
are functions of $\cos\varphi$ while
the odd coefficients vanish, $G_{2n+1}=0$. As a consequence,
the current is an even function of the flux and fulfills the Onsager symmetry.
This result also holds for finite temperatures. For instance, the linear
conductance reads,
\begin{multline}
G_0= \frac{2e^2}{h} \frac{\beta}{2\pi} \left\{ \wG\sqrt{\alpha\calt_b\calr_b}\cos(\varphi) \Im \left[\Psi_0\right]\right.\\
+ \left.\frac{\wG}{2} \left[\alpha\left(1-\calt_b\cos^2(\varphi)\right) - \calt_b\right] \Re \left[\Psi_0\right] \right\}\,,
\end{multline}
where $\beta$ is the inverse temperature and
$\Psi_0 \equiv \Psi\left(\frac{1}{2} + \frac{\beta\wG}{4\pi} + i\beta\frac{\wteps_d}{2\pi}\right)$
denotes the digamma function.\cite{abram}
Note that we have subtracted the offset term $\int d\omega~ \calt_b [f_L(\omega) - f_R(\omega)]$.
In the zero temperature case, $G_0$ becomes,
\begin{multline}\label{eq_g0nint}
G_0(\varphi)= \frac{2e^2}{h} \left\{ \calt_b -\frac{\wG\wteps_d\sqrt{\alpha\calt_b\calr_b}\cos(\varphi)}{(\wteps_d)^2+\frac{\wG^2}{4}}\right.\\
+ \left.\frac{\wG^2 \left[\alpha\left(1-\calt_b \cos^2(\varphi)\right)-\calt_b\right]}{4\left[(\wteps_d)^2+\frac{\wG^2}{4}\right]} \right\}\,,
\end{multline}
where
\beq\label{eq:varepsren}
\wteps_d = \veps_d - \frac{1}{4}\sqrt{\alpha\calt_b} \Gamma \cos(\varphi)\,.
\edq
Equation~(\ref{eq_g0nint}) is clearly an even function of $\varphi$.
These results show that in the absence of interactions,
transport is $B$-symmetric to all orders in voltage.
However, a word of caution is in order.
Neglecting interactions in the nonlinear regime of transport can lead to
unphysical results (e.g., gauge invariance can be broken).\cite{christen,comment}
Therefore, to give reliable results away from equilibrium
we must include interactions at least in the lowest
level of approximation.

\subsection{Hartree approximation and even-odd behavior}
We now introduce interactions in the most simple way, namely,
we use in Eq.~(\ref{eq:eom}) the following decoupling,
\beq
\langle\langle d_{\sm} n_{d\bsm},d_{\sm}^{\dag} \rangle\rangle
\approx \langle n_{d\bsm}\rangle\langle\langle d_{\sm},d_{\sm}^{\dag} \rangle\rangle\,.
\edq
This Hartree approximation is well known to spontaneously generate
local moment formation\cite{anderson} in the quantum dot. Although this result
is physically meaningless since the Hamiltonian [Eq.~(\ref{eq:model})] is invariant under
spin rotations, the approximation is useful as a benchmark
for more elaborate models.

The retarded Green function is found to be,
\beq\label{eq:gdsmdsmmf}
\calg_{d\sm,d\sm}^{r}(\omega) = \frac{1}{\omega - \veps_d - U\nbraket{n_{d\bsm}} + \frac{1}{4}\sqrt{\calt_b} \Gamma \cos(\varphi) + \frac{i}{2} \widetilde{\Gamma}}\,,
\edq
which was anticipated in the Introduction.
The electron occupation $\nbraket{n_{d\sm}}$ can be obtained from
\beq \label{eq:occ}
\nbraket{n_{d\sm}} = \frac{1}{2\pi i} \int d\omega~ \calg_{d\sm,d\sm}^{<}(\omega)\,.
\edq

In general, for interacting systems,
the lesser Green's function cannot be directly obtained
from the equation-of-motion technique without introducing additional assumptions.
However, we note that only the integral of $\calg_{d\sm,d\sm}^<(\omega)$
is, in fact, needed in Eq.~\eqref{eq:occ}.
This observation allows us to bypass any approximation
involved in computing $\calg_{d\sm,d\sm}^<(\omega)$, yielding \cite{Sun02}
\beq\label{eq:ndsmgr}
\nbraket{n_{d\sm}} =-\frac{1}{2\pi i} \int d\omega\,
f_{peq}(\omega) \left[\calg_{d\sm,d\sm}^{r}(\omega) - \calg_{d\sm,d\sm}^{a}(\omega)\right] \,,
\edq
An alternative derivation is presented in App.~A.
In Eq.~\eqref{eq:ndsmgr}, $f_{peq}(\omega)$
denotes a {\em pseudoequilibrium} distribution function
which is not, quite generally, of the Fermi-Dirac type,
\begin{multline}
f_{peq}(\omega) = \frac{1}{\widetilde{\Gamma}} \left\{ \frac{1}{2} \widetilde{\Gamma}\sqrt{\alpha\calt_b}\sin(\varphi) \left[f_L(\omega)-f_R(\omega)\right]\right.\\
+ \left.\frac{1}{(1+\xi)^2}\left[\left(\Gamma_L + \xi\Gamma_R\right)f_L(\omega) + \left(\Gamma_R + \xi\Gamma_L\right)f_R(\omega)\right]\right\}\,.
\label{eq:fneq}
\end{multline}

Setting $\calg_{d\sm,d\sm}^{a}(\omega)=[\calg_{d\sm,d\sm}^{r}(\omega)]^{\ast}$,
we see that Eqs.~(\ref{eq:gdsmdsmmf}), (\ref{eq:ndsmgr}) and~(\ref{eq:fneq})
form a closed system of equations which must be solved
self-consistently. But before proceeding with such a calculation,
we point out to the presence of a {\em $\varphi$-asymmetric} term
already in Eq.~(\ref{eq:fneq}). Note that this term is nonzero
only in the nonequilibrium case ($f_L\neq f_R$).

\subsubsection{Zero temperature case}

For $\Gamma_L = \Gamma_R$, the pseudoequilibrium distribution function
can be simplified,
\beq
f_{peq}(\omega)=\frac{\left[f_L(\omega) + f_R(\omega)\right] + 
\sqrt{\calt_b}\sin(\varphi) \left[f_L(\omega) - f_R(\omega)\right]}{2}\,,
\edq
and from Eq.~\eqref{eq:ndsmgr} we find an
{\em exact} expression for the occupation,
\begin{multline}\label{eq:nbraketndsm}
\nbraket{n_{d\sm}} = \frac{1}{2\pi} \left\{ \left[1 + \sqrt{\calt_b}\sin(\varphi) \right])\right.\\
\times\cot^{-1} \left[\frac{2\left(\wteps_d + U\nbraket{n_{d\bsm}} -\mu_{L}\right)}{\wG} \right]    \\
+ \left.\left[1 - \sqrt{\calt_b}\sin(\varphi) \right]\cot^{-1} \left[\frac{2\left(\wteps_d + U\nbraket{n_{d\bsm}} -\mu_{R}\right)}{\wG} \right] \right\}\,.
\end{multline}
As introduced in Sec.~\ref{sec_intro},
injection from the left lead contributes to the dot occupation
with a term $1 + \sqrt{\calt_b}\sin(\varphi)$
while the contribution from the right lead
is given by $1 - \sqrt{\calt_b}\sin(\varphi)$.
Both terms cancel out at equilibrum, regardless of interaction,
but survive in the presence of a finite bias.

We solve Eq.~\eqref{eq:nbraketndsm} iteratively. We write,
\beq\label{eq_nexpansion}
\nbraket{n_{d\sm}} = \nbraket{n_{d\sm}}^{(0)} + \nbraket{n_{d\sm}}^{(1)} (eV) + \nbraket{n_{d\sm}}^{(2)} (eV)^2 + \cdots\,,
\edq
insert this expansion in Eq.~\eqref{eq:nbraketndsm}
and assume $\nbraket{n_{d\sm}} = \nbraket{n_{d\bsm}}$.
Then, we obtain the following expansion coefficients,
\bes
\begin{align}
\nbraket{n_{d\sm}}^{(0)} &= \frac{1}{\pi} \tan^{-1} \left[ \frac{\wG}{2\wteps_d^0} \right] \,,\\
\nbraket{n_{d\sm}}^{(1)} &= \frac{\frac{\wG}{2} \sqrt{\calt_b}\sin(\varphi)}{2\left\{ \pi \left[\left(\wteps_d^0\right)^2 + \frac{\wG^2}{4} \right] + \frac{\wG}{2} U \right\}} \,,\\
\nbraket{n_{d\sm}}^{(2)} &= 
\frac{\wG\wteps_d^0 \left[1+4\left( U\nbraket{n_{d\sm}}^{(1)}\right)^2 - 4U\sqrt{\calt_b}\sin(\varphi)\nbraket{n_{d\sm}}^{(1)}\right]}
{4\left\{ \pi \left[\left(\wteps_d^0\right)^2 + \frac{\wG^2}{4} \right] + \frac{\wG}{2} U \right\} \left[\left(\wteps_d^0\right)^2 + \frac{\wG^2}{4} \right] }\,,
\end{align}
\begin{multline}
\nbraket{n_{d\sm}}^{(3)} = 3!
\left\{ 2\wG U \left[ 4U^2 \left(12(\wteps_d^0)^2 - \wG^2\right)\left(\nbraket{n_{d\sm}}^{(1)}\right)^3 
\right. \right. 
\\
\left.
+ 3\nbraket{n_{d\sm}}^{(1)}
\left[ 12(\wteps_d^0)^2 - \wG^2 - 4U\wteps_d^0 \left(4(\wteps_d^0)^2 + \wG^2\right)\nbraket{n_{d\sm}}^{(2)} \right] \right] \\
+ \sqrt{\calt_b}\sin(\varphi) \left[ \wG\left(-12(\wteps_d^0)^2 + \wG^2\right)\left(1+12\left(U\nbraket{n_{d\sm}}^{(1)}\right)^2\right)
\right.
\\
\left.\left.
+ 12\wG U\wteps_d^0 \left(4(\wteps_d^0)^2 + \wG^2\right)\nbraket{n_{d\sm}}^{(2)} \right] \right\} \\
\times
\left\{192\left[ \pi \left(\left(\wteps_d^0\right)^2 + \frac{\wG^2}{4} \right) + \frac{\wG}{2} U \right] \left(\left(\wteps_d^0\right)^2 + \frac{\wG^2}{4} \right)^2 \right\}^{-1}\,,
\end{multline}
\label{eq:ndexp}
\eds
where we have defined
\beq
\wteps_d^0 = \veps_d + U\nbraket{n_{d\sm}}^{(0)} - \frac{1}{4}\sqrt{\calt_b}\Gamma\cos(\varphi)\,,
\edq
which is an even function of the field.
>From Eq.~\eqref{eq:ndexp}, we infer the relations,
\bes\label{eq_ndsmexpall}
\begin{align} 
\nbraket{n_{d\sm}}^{(0)}(\varphi) &= +\nbraket{n_{d\sm}}^{(0)}(-\varphi) \\
\label{eq_ndsmexp}\nbraket{n_{d\sm}}^{(1)}(\varphi) &= -\nbraket{n_{d\sm}}^{(1)}(-\varphi) \\
\nbraket{n_{d\sm}}^{(2)}(\varphi) &= +\nbraket{n_{d\sm}}^{(2)}(-\varphi) \\
\label{eq_ndsmexp3}\nbraket{n_{d\sm}}^{(3)}(\varphi) &= -\nbraket{n_{d\sm}}^{(3)}(-\varphi)
\end{align}
\eds
Therefore, we can then conclude that the nonequilibrium charge response of the system
is not a symmetric function of the flux, as shown in the odd coefficients in Eq.~(\ref{eq_ndsmexp})
and Eq.~(\ref{eq_ndsmexp3}). In fact, these coefficients are antisymmetric
when the field is inverted. On the other hand,
the even coefficients are all symmetric
under reversal of $\varphi$. This restoration of the Onsager symmetry for
the even coefficients of a current--voltage expansion has been observed
experimentally in rings (see Ref.~\onlinecite{let06}). In this section,
we do not attempt to make a direct comparison with the experimental data
since the conductance within the Hartree approximation would give wrong results
due to the aforementioned breaking of the spin rotation symmetry.
Nevertheless, in the next section we demonstrate that this even-odd behavior
is not an artefact of the Hartree approximation
and persists in a better treatment of Coulomb interaction
from which a physically meaningful conductance can be extracted.

\subsubsection{Nonzero temperature case}

For completeness, we briefly discuss the nonzero temperature
case. Equation~(\ref{eq:nbraketndsm}) is replaced with
\begin{multline}\label{eq:nbraketndsm2}
\nbraket{n_{d\sm}} = 
\frac{1}{2\pi} \left\{ 
\left(1 + \sqrt{\calt_b}\sin(\varphi) \right) \right.
\\
\times
\left[ \frac{1}{2} - \frac{1}{\pi} \Im \left[ \Psi\left(\frac{1}{2} + \frac{\beta\wG}{4\pi} + i\beta\frac{(\wteps_d-\mu_{L})}{2\pi}\right) \right] \right] 
\\
+ \left(1 - \sqrt{\calt_b}\sin(\varphi) \right) 
\\
\left. \times
\left[ \frac{1}{2} - \frac{1}{\pi} \Im \left[ \Psi\left(\frac{1}{2} + \frac{\beta\wG}{4\pi} + i\beta\frac{(\wteps_d-\mu_{R})}{2\pi}\right) \right] \right] \right\}\,.
\end{multline}
We now substitute the expansion Eq.~(\ref{eq_nexpansion})
in Eq.~(\ref{eq:nbraketndsm2}) and find exactly the same relations as
Eq.~(\ref{eq_ndsmexp}).
For instance, the first two expansion coefficients read,
\begin{align}
\nbraket{n_{d\sm}}^{(0)} &= \frac{1}{2\pi} 
\left\{ 1 - \frac{2}{\pi} \Im\left[ \Psi_0 \right]\right\} \\
\nbraket{n_{d\sm}}^{(1)} &= \frac{\beta}{4\pi^3}\frac{ \sqrt{\calt_b}\sin(\varphi) 
\Re\left[ \Psi_0^{(1)} \right] }
{1 + \frac{\beta U}{2\pi^3}\Re\left[ \Psi_0^{(1)} \right]}\,,
\end{align}
where
$\Psi_0^{(n)} = \Psi^{(n)} \left(\frac{1}{2} + \frac{\beta\wG}{4\pi} + i\beta\frac{\wteps_d^0}{2\pi}\right)$
is the polygamma function (the $n$th derivative of the digamma function defined above).\cite{abram}
We again see the antisymmetric charge response of the system due to the $\sin\varphi$
term in the leading-order nonequilibrium coefficient $\nbraket{n_{d\sm}}^{(1)}$.

\section{Coulomb blockade regime}\label{sec_cb}
In the Coulomb blockade regime of two-terminal quantum dots, transport takes place only through
two resonances approximately located at $\varepsilon_d$ and $\varepsilon_d+U$.
Clearly, the retarded Green function given by Eq.~(\ref{eq:gdsmdsmmf})
does not show this behavior and consequently we must perform a higher-order
truncation in Eq.\ (\ref{eq:eom}). This way, one
obtains the equation of motion for
$\langle\langle d_{\sm} n_{d\bsm},d_{\sm}^{\dag} \rangle\rangle$:
\begin{multline}
\left( \omega - \veps_d - U \right) \langle\langle d_{\sm}n_{d\bsm},d_{\sm}^{\dag} \rangle\rangle = \langle n_{d\bsm} \rangle \\
+ \sum_{\alpha,k} V_{\alpha}^{\ast}  \langle\langle c_{\alpha k\sm} n_{d\bsm}, d_{\sm}^{\dag} \rangle\rangle
+ \sum_{\alpha,k} V_{\alpha}^{\ast} \langle\langle d_{\bsm}^{\dag} c_{\alpha k \bsm} d_{\sm},d_{\sm}^{\dag} \rangle\rangle \\
- \sum_{\alpha, k} V_{\alpha} \langle\langle c_{\alpha k \bsm}^{\dag} d_{\bsm}d_{\sm},d_{\sm}^{\dag} \rangle\rangle \,.
\label{eq:dsam}
\end{multline}
To obtain the two-peak solution, we keep only the first term
on the right-hand side of Eq.~\eqref{eq:dsam}, calculate its equation of motion,
and make the following approximations:\cite{Hewson66}
\bes
\begin{align}
\langle\langle c_{\alpha k \sm} c_{\alpha'k'\bsm}^{\dag} d_{\bsm}, d_{\sm}^{\dag} \rangle\rangle &\approx 0 \,,\\
\langle\langle c_{\alpha k \sm} d_{\bsm}^{\dag} c_{\alpha'k'\bsm}, d_{\sm}^{\dag} \rangle\rangle &\approx 0\,.
\end{align}
\label{eq:hfdec}
\eds
Then,
\begin{multline}
\calg_{d\sm,d\sm}^{r}(\omega) = \frac{1-\nbraket{n_{d\bsm}}}{\omega - \veps_d + \frac{1}{4}\sqrt{\alpha\calt_b}\Gamma\cos(\varphi) + \frac{i}{2}\wG}\\
+ \frac{\nbraket{n_{d\bsm}}}{\omega - \veps_d - U + \frac{1}{4}\sqrt{\alpha\calt_b}\Gamma\cos(\varphi) + \frac{i}{2}\wG}\,.
\end{multline}
We note that the retarded Green function now correctly
shows two peaks located at $\veps_d-\frac{1}{4}\sqrt{\alpha\calt_b}\Gamma\cos(\varphi)$ 
and $\veps_d-\frac{1}{4}\sqrt{\alpha\calt_b}\Gamma\cos(\varphi)+ U$ with weights $1-\nbraket{n_{d\bsm}}$ and $\nbraket{n_{d\bsm}}$, respectively.

Using Eq.~\eqref{eq:ndsmgr} we find that the occupation is given by
\beq
\nbraket{n_{d\sm}} = \frac{\tau_L\mathcal{I}_L(\wteps_d^0) +\tau_R\mathcal{I}_R(\wteps_d^0)}
{4 + \tau_L \left(\mathcal{I}_L(\wteps_d^0) - \mathcal{I}_L(\wteps_d^U)\right) 
+ \tau_R \left(\mathcal{I}_R(\wteps_d^0) - \mathcal{I}_R(\wteps_d^U)\right)}\,,
\edq
where we have used the following definitions,
\bes
\begin{align}
\tau_{L(R)}& =\frac{2(\Gamma_{L(R)}+\xi\Gamma_{R(L)})}{(1+\xi)\Gamma} \pm \sqrt{\alpha\calt_b}\sin(\varphi) \,,\\
\mathcal{I}_{\alpha}(x)& = 1 - \frac{2}{\pi}\Im\left[\Psi\left(\frac{1}{2} + \frac{\beta\wG}{4\pi} + i\frac{\beta(x-\mu_{\alpha})}{2\pi}\right) \right]\,, \\
\wteps_d^0 &= \veps_d - \frac{1}{4} \sqrt{\alpha\calt_b}\Gamma\cos(\varphi) \,, \\
\wteps_d^U &= \veps_d + U - \frac{1}{4} \sqrt{\alpha\calt_b}\Gamma\cos(\varphi) \,.
\end{align}
\eds

For symmetric couplings
($\Gamma_L = \Gamma_R$) or a completely open nonresonant channel ($\xi = 1$),
we find for the particular case of symmetric bias ($\mu_L = -\mu_R \equiv V/2$) that
\beq
\nbraket{n_{d\sm}}(\varphi,V) = \nbraket{n_{d\sm}}(-\varphi,-V) \,.
\label{eq:eosym}
\edq
Physically, this corresponds to an invariance of the whole system
when {\em both} the magnetic field and the electric bias are inverted.\cite{kon02}
A series expansion in $V$ then yields,
\begin{multline}
\nbraket{n_{d\sm}}^{(0)}(\varphi) + \nbraket{n_{d\sm}}^{(1)}(\varphi) V + \cdots
= 
\\
\nbraket{n_{d\sm}}^{(0)}(-\varphi) - \nbraket{n_{d\sm}}^{(1)}(-\varphi) V + \cdots .
\end{multline}
Thus, we have
\bes
\begin{align}
\nbraket{n_{d\sm}}^{(2n)}(\varphi) &= +\nbraket{n_{d\sm}}^{(2n)}(-\varphi) \,, \\
\nbraket{n_{d\sm}}^{(2n+1)}(\varphi) &= -\nbraket{n_{d\sm}}^{(2n+1)}(-\varphi) \,.
\end{align}
\eds
These equations represent a generalization of the Hartree case
[Eqs.~\eqref{eq_ndsmexpall}] to the Coulomb blockade regime.
It then follows that
\beq
\calt_{\sm}(\varphi,V) = \calt_{\sm}(-\varphi,-V)\,,
\edq
and
\beq
I(\varphi,V) = -I(-\varphi,-V)\,.
\edq
A further expansion of $I$ in powers of $V$ finally gives,
\bes
\begin{align}
G_{2n}(\varphi) &= G_{2n}(-\varphi) \,, \\
\label{eq:condodd}G_{2n+1}(\varphi) &= -G_{2n+1}(-\varphi) \,,
\end{align}
\label{eq:condeo}
\eds
i.e., the even (odd) conductance coefficients are symmetric
(antisymmetric) functions of the flux. 
We see here a crucial difference compared to the noninteracting case discussed earlier.
For $U=0$ the current is always a symmetric function of $\varphi$ regardless
of the applied voltage. In the interacting case, we find that the odd
coefficients of the conductance are not invariant when the field orientation
is inverted.

The above property
is not general and can be traced back to the spatial symmetry
of the system (symmetric couplings and symmmetric bias).
In the more general case of asymmetric couplings ($\Gamma_L \ne \Gamma_R$),
the occupation symmetry given by Eq.~\eqref{eq:eosym} is not fulfilled.
To see this, for simplicity we take the limit $U\to\infty$. 
Then, the occupation is given by
\beq
\nbraket{n_{d\sm}} = \frac{\mathcal{I}_L + \mathcal{I}_R
+ A(\varphi)\left(\mathcal{I}_L - \mathcal{I}_R\right)}
{4 + \mathcal{I}_L + \mathcal{I}_R 
+ [\sqrt{\calr_b}\delta\Gamma + \sqrt{\alpha\calt_b}\sin\varphi]
(\mathcal{I}_L - \mathcal{I}_R)}\,,
\label{eq:nexpinf}
\edq
where $\mathcal{I}_\alpha$ is evaluated at $x=\wteps_d^0$ and
we have defined
\beq
A(\varphi) = \left[\sqrt{\calr_b}\delta\Gamma + \sqrt{\alpha\calt_b}\sin(\varphi)\right]\,,
\label{eq:aphi}
\edq
and
\beq
\delta\Gamma = \frac{\Gamma_L - \Gamma_R}{\Gamma}\,.
\edq
We now expand the occupation as a function of $V$ and find that the expansion coefficients are given by
\bes
\begin{align}
\nbraket{n_{d\sm}}^{(2n)} &= \sum_{m=0}^{n} C_{2m}^{(2n)}(\varphi) A^{2m}(\varphi) \,,
\\
\nbraket{n_{d\sm}}^{(2n+1)} &= \sum_{m=0}^{n} C_{2m+1}^{(2n+1)}(\varphi) A^{2m+1}(\varphi)\,,
\end{align}
\eds
with the $C$'s fulfilling,
\beq
C_{m}^{(n)}(\varphi) = C_{m}^{(n)}(-\varphi)\,.
\edq
We give the explicit expressions for the first leading-order coefficients, 
\bes
\begin{align}
\nbraket{n_{d\sm}}^{(0)} &= \frac{\pi - 2\Im\Psi_0}{3\pi - 2\Im\Psi_0}\,,
\\
\nbraket{n_{d\sm}}^{(1)} &= \frac{\Re\Psi_0^{(1)}A(\varphi)}
{\left(3\pi - 2\Im\Psi_0\right)^2/\beta}\,,
\\
\nbraket{n_{d\sm}}^{(2)} &= \frac{(3\pi - 2\Im \Psi_0)\Im \Psi_0^{(2)}
-4 \left[\Re\Psi_0^{(1)}\right]^2
A^2(\varphi)}
{8\pi \left(3\pi - 2\Im\Psi_0\right)^3/\beta^2}\,.
\end{align}
\eds

\begin{figure}
\centering
\includegraphics[width=0.4\textwidth]{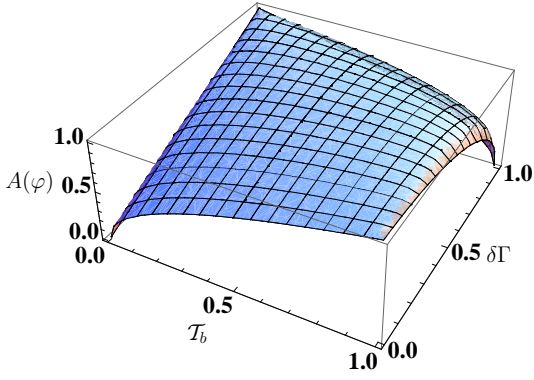}
\caption{(Color online) $A(\varphi)$ as a function of $\calt_b$ and $\delta\Gamma$ for $\varphi = \pi/2$. Refer to Eq.~\eqref{eq:aphi}.}
\label{fig:aphi}
\end{figure}

\begin{figure}
\centering
\includegraphics[width=0.45\textwidth]{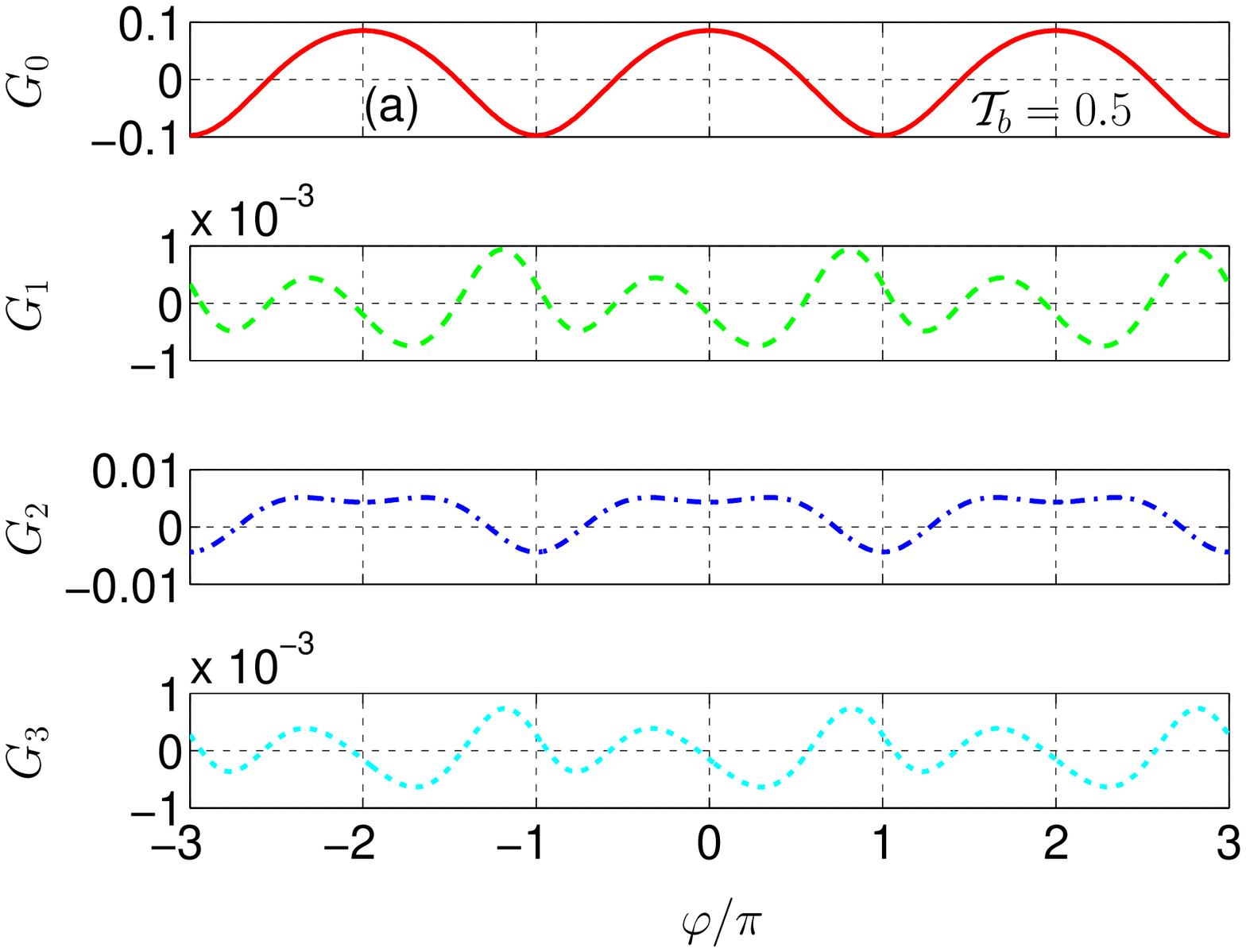}
\includegraphics[width=0.45\textwidth]{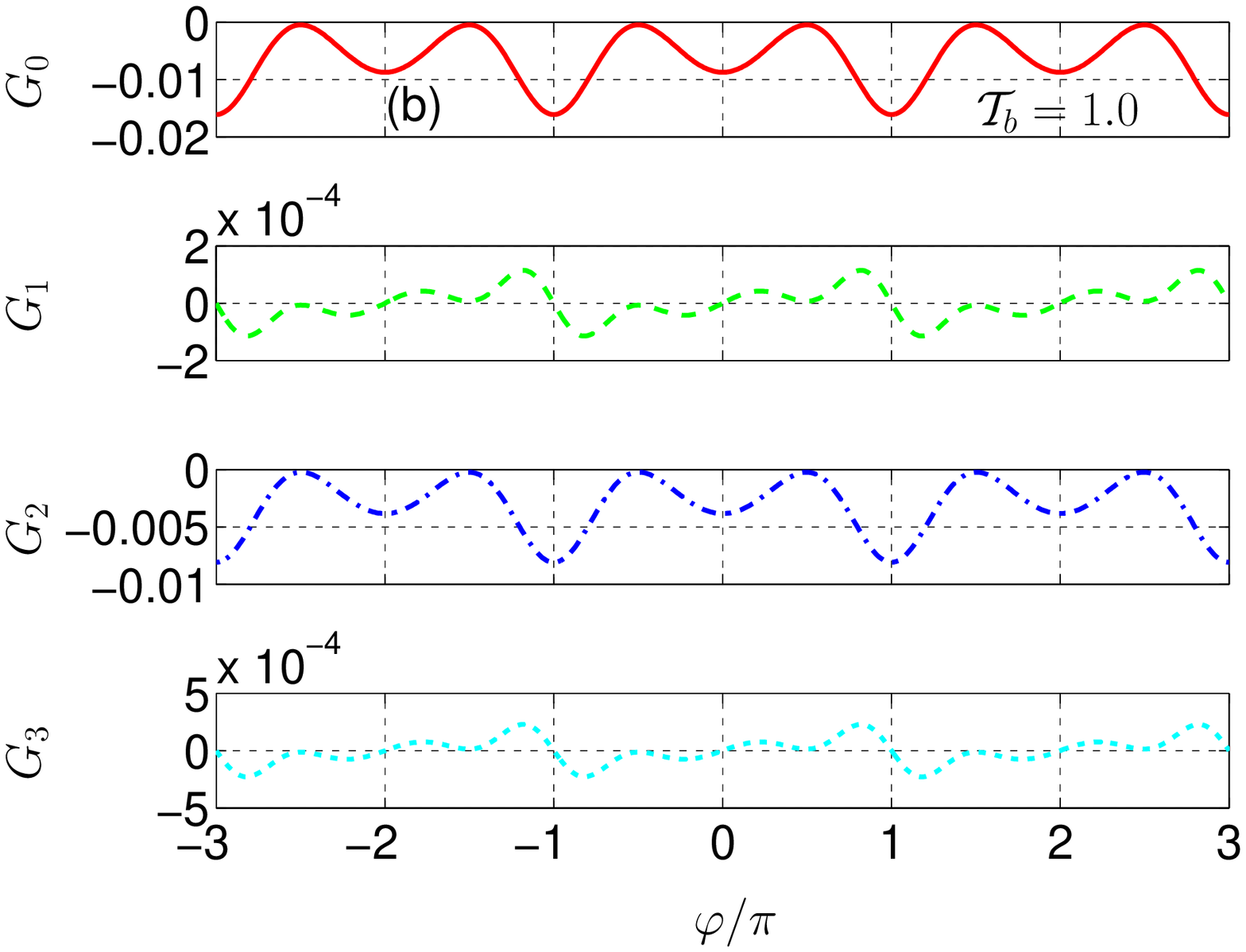}
\caption{(Color online) Conductance coefficients as a function of $\varphi$.
$G_0$ and $G_2$ are offset for simplicity.
Parameters are $\Gamma_0 = 1$, $\delta\Gamma = 0.2$, $\Gamma_{\alpha} = \Gamma_0(1 \pm \delta\Gamma)$, 
$\veps_d = -5\Gamma_0$, and $T = \Gamma_0$.
Here, the conductance coefficients are scaled by $e^{n+2}/(2^{n-1}n!h)$.}
\label{fig:conductancecoeff}
\end{figure}

Now, the dimensionless function $A(\varphi)$
is a small quantity for almost all cases. We show in Fig.~\ref{fig:aphi}
that $A$ is always smaller than 1 even for the case $\varphi=\pi/2$.
As a result, we can safely neglect $A^n(\varphi)$ for all $n> 1$,
thus keeping the first-order term only.
This implies that the even expansion coefficients are always symmetric
under the reversal of $\varphi$ but the odd ones do not show any particular symmetry
since $\Gamma_L\ne \Gamma_R$.
Since we know that the transmission $\calt_{\sm}(\omega)$ from Eq.~\eqref{eq:current}
obeys the same symmetry as the occupation $\nbraket{n_{d\sm}}$,
it follows that
\bes
\begin{align}
G_{2n}(\varphi) &= G_{2n}(-\varphi) \,, \\
G_{2n+1}(\varphi) &\ne -G_{2n+1}(-\varphi) \,.
\end{align}
\eds
This is precisely the behavior that was observed in Ref.~\onlinecite{let06}
for an asymmetric ring. In
Fig.~\ref{fig:conductancecoeff}
we numerically calculate the first four conductance coefficients
in the current expansion for two different values of the background
transmission. In both cases, $G_0$ obeys reciprocity, as expected.
For a partially open direct channel ($\mathcal{T}_b=0.5$),
the leading-order nonlinearity, $G_1$, is magnetoasymmetric
but the Onsager symmetry is recovered for $G_2$ and later
destroyed again in $G_3$. These results are in agreement with
the experiment.\cite{let06} In the case of a fully open
direct channel ($\mathcal{T}_b=1$) the odd coefficients
are still asymmetric but they are now odd functions
of the magnetic flux, in agreement with Eq.~\eqref{eq:condodd}.

\section{Kondo corrrelations}\label{sec_kondo}
We can now go to next order in the equation-of-motion technique
to describe the onset of Kondo correlations.
Thus, we obtain the equations of motion for the three functions appearing on the
right-hand side of Eq.~\eqref{eq:dsam}, and 
approximate the new Green's functions that appear in the procedure
by making the decouplings,~\cite{lac81}
\bes
\begin{align}
\dbraket{c_{\alpha k\sm}c_{\beta r\bsm}^{\dag}d_{\bsm},d_{\sm}^{\dag}} &\approx \nbraket{c_{\beta r\bsm}^{\dag}d_{\bsm}}\dbraket{c_{\alpha k\sm},d_{\sm}^{\dag}}\,, \\
\dbraket{c_{\alpha k\sm}d_{\bsm}^{\dag} c_{\beta r\bsm},d_{\sm}^{\dag}} &\approx \nbraket{d_{\bsm}^{\dag}c_{\beta r\bsm}}\dbraket{c_{\alpha k\sm},d_{\sm}^{\dag}}\,, \\
\dbraket{d_{\bsm}^{\dag}c_{\alpha k\bsm}c_{\beta r\sm},d_{\sm}^{\dag}} &\approx \nbraket{d_{\bsm}^{\dag}c_{\alpha k\bsm}}\dbraket{c_{\beta r\sm},d_{\sm}^{\dag}}\,, \\
\dbraket{c_{\beta r\bsm}^{\dag}c_{\alpha k\bsm}d_{\sm},d_{\sm}^{\dag}} &\approx \nbraket{c_{\beta r\bsm}^{\dag}c_{\alpha k\bsm}}\dbraket{d_{\sm},d_{\sm}^{\dag}}\,, \\
\dbraket{c_{\alpha k\bsm}^{\dag}c_{\beta r\bsm}d_{\sm},d_{\sm}^{\dag}} &\approx \nbraket{c_{\alpha k\bsm}^{\dag}c_{\beta r\bsm}}\dbraket{d_{\sm},d_{\sm}^{\dag}}\,, \\
\dbraket{c_{\alpha k\bsm}^{\dag}d_{\bsm}c_{\beta r\sm},d_{\sm}^{\dag}} &\approx \nbraket{c_{\alpha k\bsm}^{\dag}d_{\bsm}}\dbraket{c_{\beta r\sm},d_{\sm}^{\dag}}\,.
\end{align}
\eds

In what follows, we take the limit $U \to \infty$, in which case
the term $\langle\langle c_{\alpha k\bsm}^{\dag} d_{\sm} d_{\bsm},d_{\sm}^{\dag} \rangle\rangle$
does not give any contribution. After little algebra, we find,
\beq
\dbraket{d_{\sm},d_{\sm}^{\dag}} ^r
= \frac{1 - \nbraket{n_{d\bsm}} - \delta n_{d\bsm}(\omega)}{\omega + i0^+ - \veps_d - \left(1 - \delta n_{d\bsm}(\omega)\right)\Sigma_0(\omega) - \Sigma_1(\omega)} \,,
\label{eq:greensc}
\edq
where
\beq
\delta n_{d\bsm}(\omega)
= -\frac{\wG}{2\pi} \int d\omega'~ \frac{f_{peq}(\omega')}{\omega' - \omega - i0^+} \left[\dbraket{d_{\bsm},d_{\bsm}^{\dag}}_{\omega'}^r\right]^{\ast} \,,
\label{eq:deltan}
\edq
and
\bes
\begin{align}
\Sigma_0(\omega) &= -i\frac{\wG}{2} - \frac{1}{4}\sqrt{\alpha\calt_b}\Gamma\cos(\varphi) \,,
\\
\Sigma_1(\omega) 
&= -\frac{\wG}{2\pi} \int d\omega'~ \frac{f_{peq}(\omega')}{\omega' - \omega - i0^+} \nonumber \\
&\times \left\{ 1 + \left[\Sigma_0(\omega')\dbraket{d_{\bsm},d_{\bsm}^{\dag}}_{\omega'}^r\right]^{\ast}\right\} \,.
\end{align}
\eds
The derivation of this expression for the retarded Green function is explained in App.~\ref{app:expeval}.

To lowest order in $\Gamma$, Eq.~\eqref{eq:greensc} can be further simplified as
\beq
\dbraket{d_{\sm},d_{\sm}^{\dag}} 
= \frac{1 - \nbraket{n_{d\bsm}}}{\omega + i0^+ - \veps_d - \Sigma_r(\omega)} \,,
\label{eq:greenscsim}
\edq
where
\beq
\Sigma_r(\omega) = \Sigma_0(\omega) + \Sigma_{1}(\omega) \,,
\edq
with
\beq
\Sigma_{1}(\omega) 
= -\frac{\wG}{2\pi} \int d\omega'~ \frac{f_{peq}(\omega')}{\omega' - \omega - i0^+} \,.
\edq
Here, we note that for $\Gamma_L = \Gamma_R$ the self-energy obeys the following symmetry
\beq
\Sigma_r(\varphi,V) = \Sigma_r(-\varphi,-V) \,.
\edq
In turn, this property implies that the occupation and the conductance
obey the even-odd symmetry
also in the Kondo regime (at least when the coupling is not very strong).

Together with Eqs.\ (\ref{eq:occ}) and~(\ref{eq:fneq}) the Green function
of Eq.~\eqref{eq:greensc} can be obtained from a self-consistent procedure.
But before solving this system of equations using numerical methods, we briefly discuss two limits
(high and low temperatures) to clarify the origin of magnetoasymmetries in the Kondo regime.

\subsection{High-temperature regime}

In this case, $\delta n_{d\bsm}$ is a small correction and an expansion can be done.
To first order in $\Gamma$ 
it can be shown that the position of the virtual level is renormalized to $\veps_d'$
\beq
\veps_d' = \veps_d - \frac{1}{4}\sqrt{\alpha\calt_b}\Gamma\cos(\varphi) 
- \frac{\wG}{2\pi} \ln \frac{\sqrt{(\veps_d' - E_F)^2 + (\pi/\beta)^2}}{D} \,,
\edq
for $\beta(\omega-E_F)\gg 1$.
If $\beta |\veps_d - E_F| \gg 1$ we have
\beq
\veps_d' = \veps_d - \frac{1}{4}\sqrt{\alpha\calt_b}\Gamma\cos(\varphi) 
- \frac{\wG}{2\pi} \ln \frac{|\veps_d' - E_F|}{D} \,.
\edq
From the equation above, the Kondo temperature is given by
\beq
k_B T_K = D \exp\left[\frac{2\pi}{\wG}\left(\veps_d - \frac{1}{4}\sqrt{\alpha\calt_b}\Gamma\cos(\varphi)\right)\right] \,.
\label{eq:kondotemp}
\edq
The Kondo temperature marks the energy scale below which nontrivial spin fluctuations
start to play a dominant role, leading to an antiferromagnetic exchange between
the dot electron and the conduction electrons. We note that for a quantum dot
inserted in an Aharonov-Bohm ring and in the presence of an applied flux, $T_K$
depends on $\xi$ and $\varphi$ but the dependence on the flux is weak.\cite{lop05}
Furthermore, the Kondo temperature is a static quantity and, as such,
is always a symmetric function of the flux.

\begin{figure}
\centering
\includegraphics[width=0.45\textwidth]{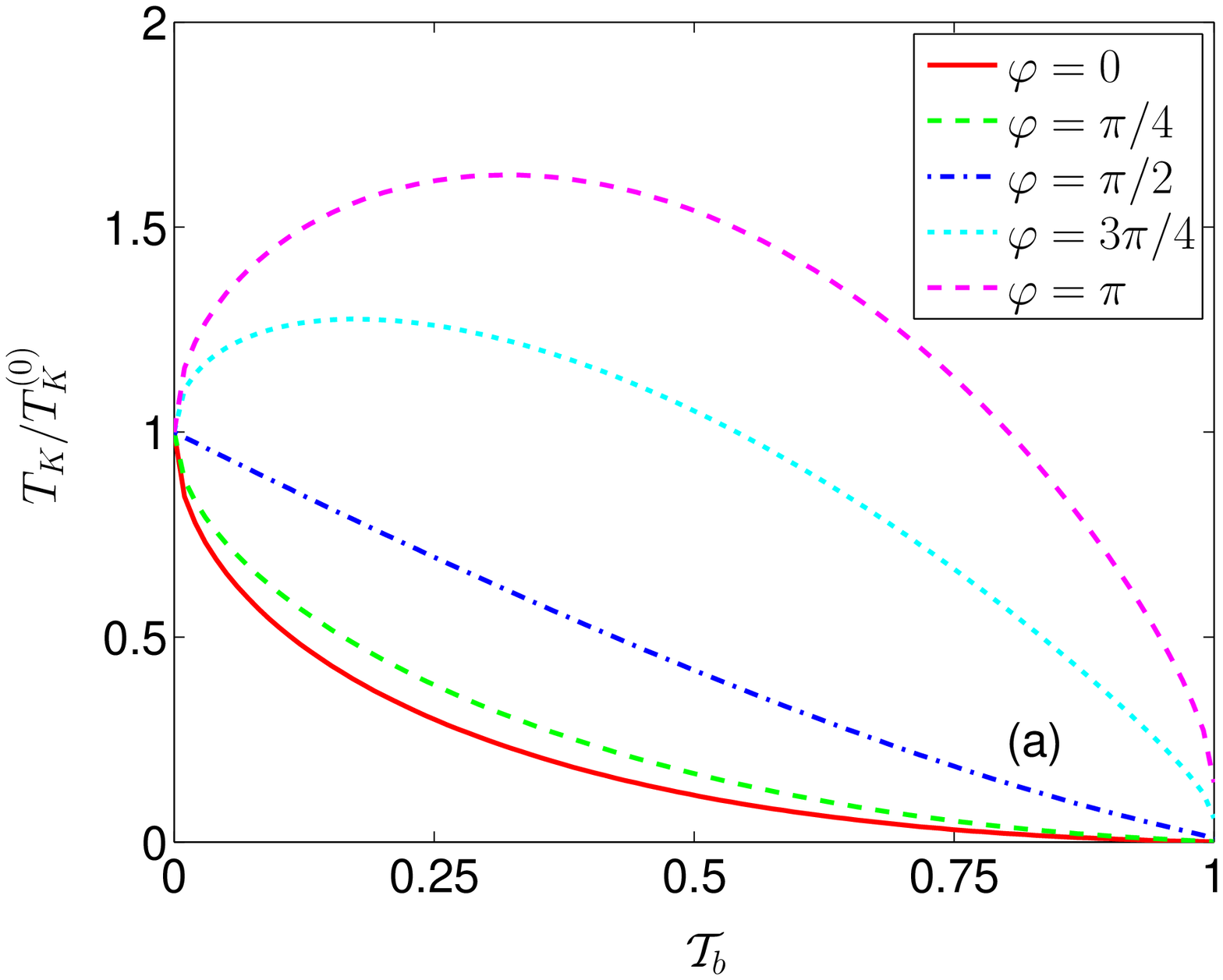}
\includegraphics[width=0.45\textwidth]{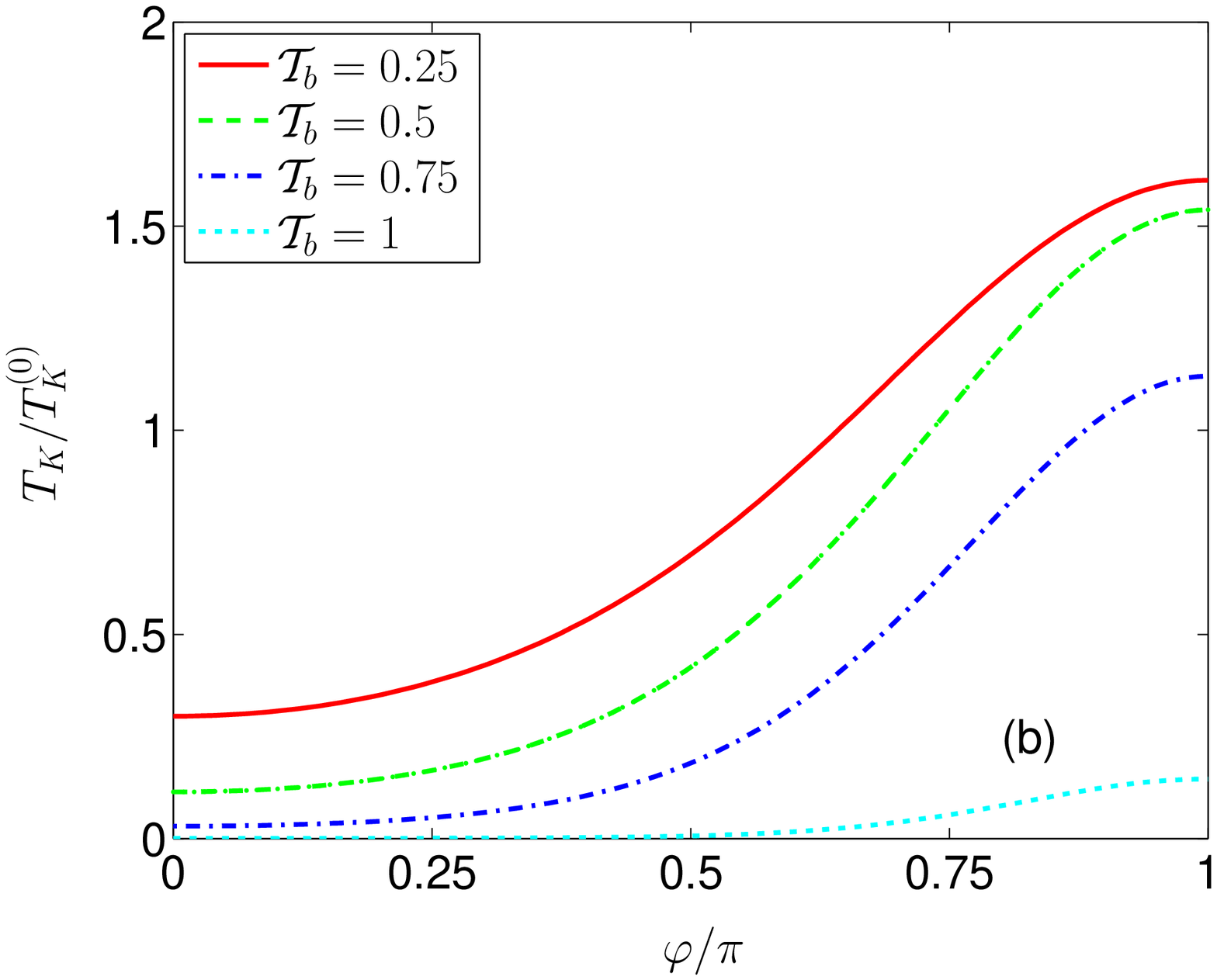}
\caption{(Color online) Kondo temperature $T_K$ as a function of $\calt_b$ and $\varphi$.
The parameters are $D=1$, $\veps_d = -0.05$, $U = \infty$, and $\Gamma_L = \Gamma_R = 0.031$.
For these parameters, $T_K^{(0)} \approx 0.0063$.
}
\label{fig:eomTK}
\end{figure}

We compare in Fig. \ref{fig:eomTK}
the value of $T_K$ with the Kondo temperature of a two-terminal quantum dot,
\begin{equation}
k_B T_K^{(0)} \sim D \exp\left\{ \frac{2\pi \veps_d}{\Gamma}\right\} \,,
\end{equation}
and plot the ratio $T_K/T_K^{(0)}$ as a function of the direct channel tunneling
probability [Fig. \ref{fig:eomTK}(a)] and the flux [Fig. \ref{fig:eomTK}(b)].
For $\varphi=0$ and $\calt_b=0$ we recover $T_K=T_K^{(0)}$ as expected.
As $\calt_b$ increases for $\varphi=0$, the Kondo temperature decreases
since electrons preferably travel along the upper arm. However,
for fluxes above $\varphi=\pi/2$, the level renormalization
due to $\cos\varphi$ is positive and the curve $T_K/T_K^{(0)}$
becomes nonmonotonous due to the competition
between the renormalized dot level and broadening.
This can be more clearly seen
in Fig. \ref{fig:eomTK}(b).

\subsection{Low-temperature regime}

At low-temperature $\delta n_{d\bsm}$ must be large, especially near the Fermi level.
If we suppose that $\calg_{d\sm,d\sm}^{r}(\omega)$ varies smoothly near the Fermi level,
Eq.~\eqref{eq:deltan} can be approximated as
\begin{multline}
\delta n_{d\bsm}(\omega) \approx -\frac{\wG}{2\pi} \left[\calg_{d\sm,d\sm}^{r}\right]^{\ast} 
\left[ \frac{i\pi}{2} + \ln \frac{2\pi}{\beta D} \right.
\\
\left. + \Psi\left(\frac{1}{2} - i\beta \frac{(\omega - E_F)}{2\pi}\right)\right] \,.
\label{eq:deltanapp}
\end{multline}

Inserting Eq.~\eqref{eq:deltanapp} into Eq.~\eqref{eq:greensc},
we find
\beq
\calg_{d\sm,d\sm}^{r}(E_F) = \frac{2}{\wG}\sin(\theta)e^{-i\theta} \,.
\edq
Here, the value of $\theta$ is related to the number of $d$ electrons
according to the Friedel-Langreth sum rule,\cite{Langreth} 
\beq
\theta = \frac{\pi\nbraket{n_d}}{2} \approx \frac{\pi}{2} \,.
\edq
This implies
\beq
\dbraket{d_{\sm},d_{\sm}^{\dag}}_{E_F} = \frac{1}{i\wG/2} \,.
\label{eq:GrEF}
\edq
Then, the linear conductance can be written as
\beq
G^{(0)}(\varphi)= \frac{2e^2}{h}\alpha \left[1 - \calt_b \cos^2(\varphi) \right] \,.
\label{eq:linearconductanceeom}
\edq
This result is exact in the limit $k_B T, V\to 0$. For nonzero voltages, one should take
into account that the imaginary part of
the interaction self-energy of  $\calg_{d\sm,d\sm}^{r}$ depends on $V$
but this dependence is weak for $eV\ll k_B T_K$ and can be safely neglected.
As a result, deep in the Kondo regime the conductance preserves the Onsager symmetry
since in the Fermi liquid picture the Kondo resonance behaves as a noninteracting
system with renormalized parameters. Charge fluctuations are quenched
and transport becomes $B$-symmetric.
This regime is beyond the scope of our method
and we prefer not to present numerical results for very low
temperatures. However, the expected scenario would be as follows:
for very low temperatures the current would be $B$-symmetric
and asymmetries would arise as temperature approaches $T_K^{(0)}$.
In the opposite case, for temperatures much larger than  $T_K^{(0)}$
transport is thermally assisted and the magnetoasymmetric effect
also disappears.\cite{san05} Therefore, we expect a large magnetoasymmetry
for temperatures of the order of $T_K^{(0)}$ for which charge fluctuations
are large. We confirm this expectation in the numerical results
reported below.


\subsection{Numerical results}\label{sec_num}
\begin{figure}
\centering
\includegraphics[width=0.45\textwidth]{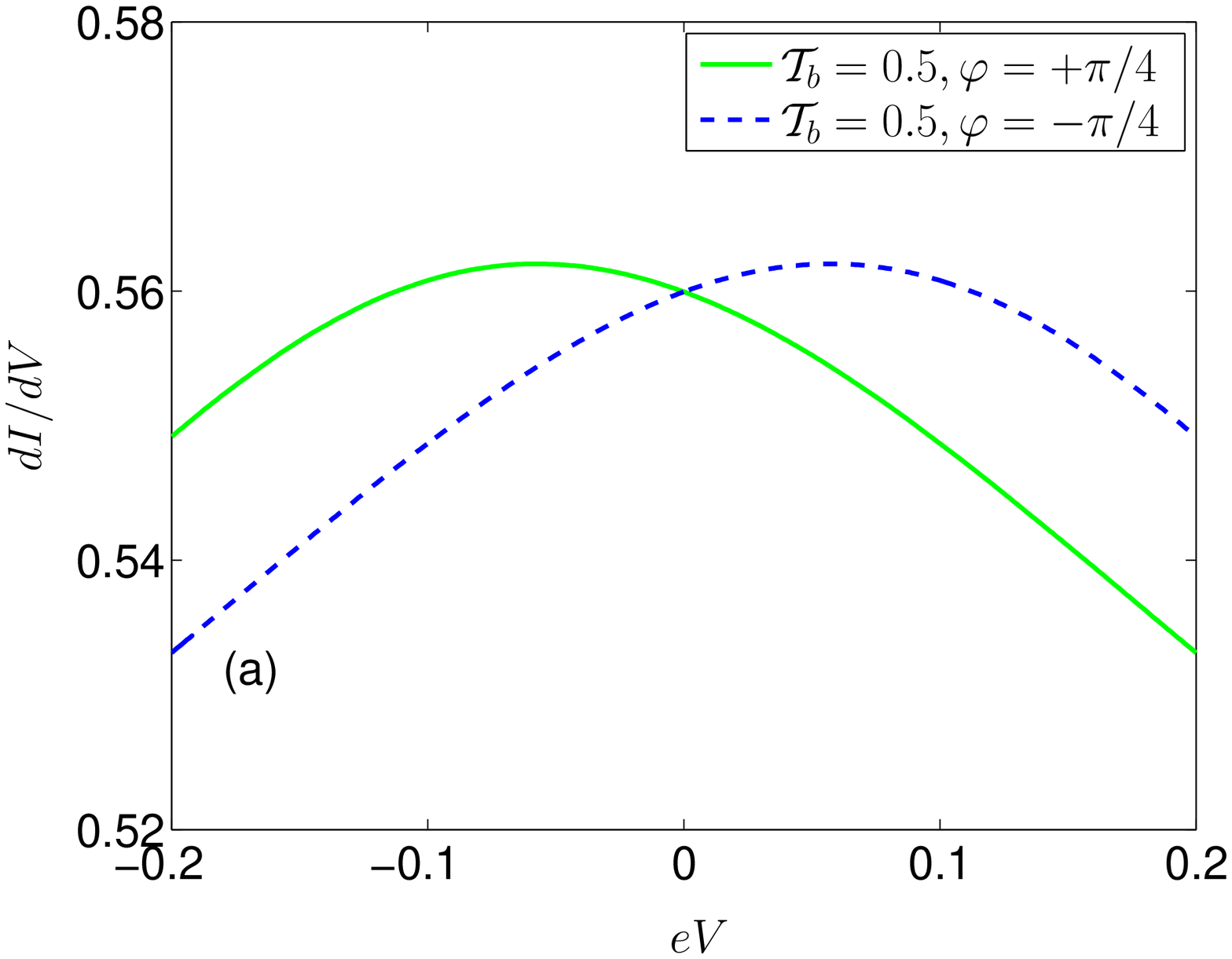}
\includegraphics[width=0.45\textwidth]{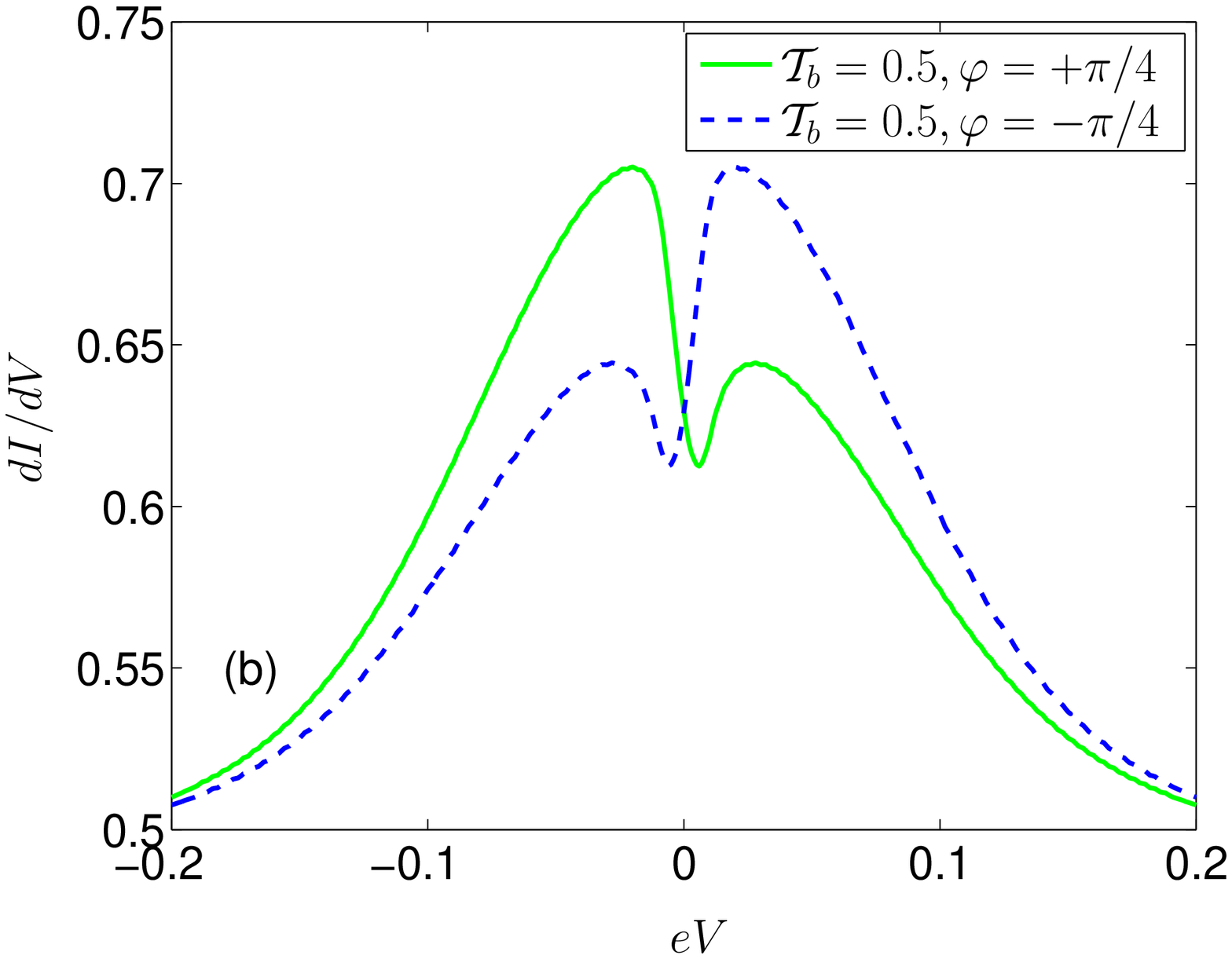}
\caption{(Color online) Differential conductance $dI/dV$ ($2e^2/h$) as a function of the applied bias $eV$
for $D = 1$, $\veps_d = -0.05$, $U = \infty$, $\Gamma_L = \Gamma_R = 0.031$, and $\varphi = \pi/4$. 
Temperatures are $T = 10T_K^{(0)}$ (a) and $0.1T_K^{(0)}$ (b).}
\label{fig:diffconductance_new}
\end{figure}

We numerically investigate the evolution of the magnetoasymmetry
when temperature is lowered from the Coulomb blockade regime
to the Kondo temperature. We first illustrate the generic behavior
in Fig.~\ref{fig:diffconductance_new},
which shows the differential condutance $dI/dV$ a a function of the applied
voltage $V$ for opposite orientations of the magnetic field.
To compute the derivative of the current we have employed a
numerical finite difference method.

In the top panel of Fig.~\ref{fig:diffconductance_new},
the temperature is large enough that Kondo correlations
can be neglected. Then, the dot is in the Coulomb blockade regime
and a small current is expected since the dot level is below the Fermi
energy ($\veps_d = -0.05$). However, the bridge channel is partially
open and the system conductance reaches around $0.55\times 2e^2/h$
at $V=0$. This value is independent of the magnetic orientation, as expected.
But when $V$ departs from equilibrium, the differential conductance
behaves differently for $+\varphi$ and $-\varphi$. We recall
that the effective position of the effective resonance depends
on the {\em charge state} of the dot, as discussed in Sec.~\ref{sec_cb}.
Since the charge response of the
system is not a symmetric function of $\varphi$, $dI/dV$ peaks
at different voltages for opposite magnetic fields.

In Fig.~\ref{fig:diffconductance_new}(b) we depict
$dI/dV$ in the low temperature case. We observe for both field orientations
a dip around $V=0$. This dip is known to arise from the destructive
interference between partial waves propagating through the upper arm
and resonantly hopping across the dot.\cite{bul01} We emphasize that
the dot bare level, $\varepsilon_d$, is the same for both calculations but in the Kondo
regime transport is dominated by the narrow resonance formed at the Fermi
level due to the higher-order tunneling processes that originate the Kondo effect.
As a result, the Fano interference between the Kondo resonance and the
background channel gives rise to the pronounced dip at zero bias.
In our case, we obtain an {\em asymmetric} lineshape
for the dip due to the magnetoasymmetric response of the dot
away from equilibrium. As a consequence, the difference in $dI/dV$ 
for $+\varphi$ and $-\varphi$ is more visible in the Kondo regime,
as can be seen in Fig.~\ref{fig:diffconductance_new}(b) compared
to Fig.~\ref{fig:diffconductance_new}(a).

\begin{figure}
\centering
\includegraphics[width=0.45\textwidth]{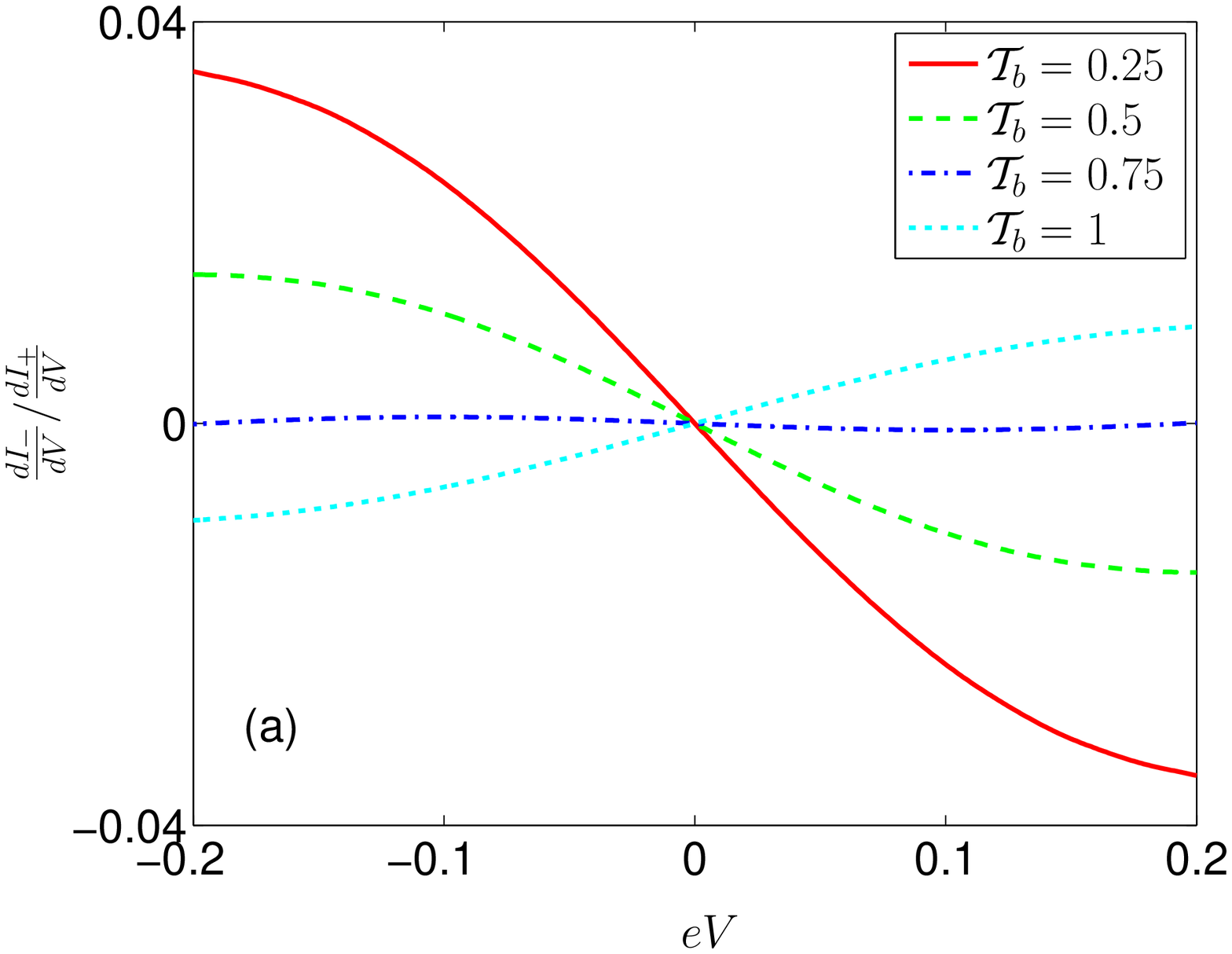}
\includegraphics[width=0.45\textwidth]{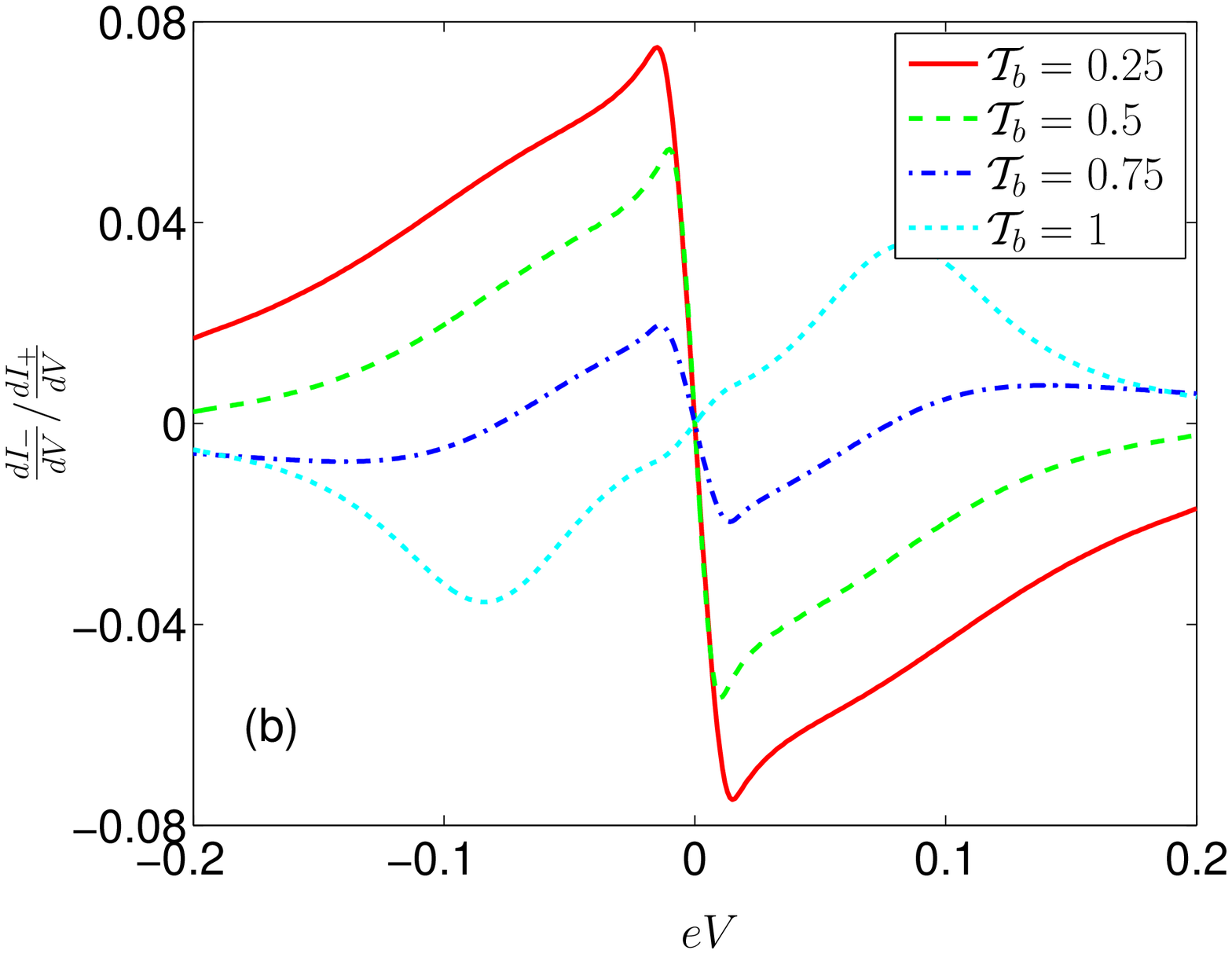}
\caption{(Color online) Ratio of differential conductances $dI_{\pm}/dV$ ($2e^2/h$) as a function of the applied bias $eV$.
Here, $dI_{\pm}/dV$ is defined as $d(I(\varphi) \pm I(-\varphi))/dV$.
Parameters are $D = 1$, $\veps_d = -0.05$, $U = \infty$, $\Gamma_L = \Gamma_R = 0.031$, and $\varphi = \pi/4$. 
Temperatures are $T = 10T_K^{(0)}$ (a) and $0.1T_K^{(0)}$ (b).}
\label{fig:diffconductance}
\end{figure}

We now define the symmetric ($+$) and antisymmetric parts ($-$)
of $dI/dV$,
\beq
\frac{dI_\pm}{dV}=\frac{dI(\varphi)}{dV}\pm\frac{dI(-\varphi)}{dV}\,.
\edq
In Fig.~\ref{fig:diffconductance} we plot the ratio between these two
components as a function of the applied bias $V$ for a fixed value of
the flux ($\varphi=\pi/4$) and for different values
of the nonresonant transmission $\mathcal{T}_b$.
In Fig~\ref{fig:diffconductance}(a)
we set the temperature to a high value
compared to the Kondo temperature. For $\mathcal{T}_b=0.5$
the magnetoasymmetry is always finite for $V\neq 0$.
For voltages around zero, the magnetoasymmetry is a linear
function of $V$ since the largest contribution stems from
the $G_1$ coefficient in the current--voltage expansion.
In the limit of high bias, the magnetoasymmetry saturates.\cite{san08}
Interestingly, with increasing $\mathcal{T}_b$
the magnetoasymmetry is reduced and changes sign
for a fixed $V$. Thefore, the sign of the asymmetry
can be {\em tuned} with the background transmission
of the nonresonant channel. The situation is similar to the
magnetoasymmetry of a two-terminal quantum dot when transport
is dominated by elastic cotunneling processes.\cite{san05}
In that case, the sign of the asymmetry can be changed with
the gate voltage which moves the dot level position
above and below the particle-hole symmetric point.\cite{san05}
In our case, $\mathcal{T}_b$ acts as an effective gate
which changes the position
of the level since $\varepsilon_0$ is renormalized
according to Eq.~(\ref{eq:varepsren}).

When temperature is lowered,
we observe that the transition from positive to negative asymmetries
as $V$ is tuned, is rather abrupt, see Fig~\ref{fig:diffconductance}(b).
We note that as voltage approaches $V=0$ one sweeps along
the strongly asymmetric dip structure found in Fig~\ref{fig:diffconductance_new}(b),
for which the difference between the cases $+\varphi$ and $-\varphi$
is most clear. Then, it is the Kondo resonance that produces
the abrupt change in the magnetoasymmetry profile as compared
to the high temperature case [Fig~\ref{fig:diffconductance}(a)].
As a consequence, Kondo correlations enhance the deviations
from the Onsager symmetry since the narrow resonance is more sensitive
to changes in the orientation of the magnetic field.\cite{zeeman}
For instance,
we observe in Fig~\ref{fig:diffconductance}(b) a revival of the
magnetoasymmetry for $\mathcal{T}_b=0.75$, which
almost vanished in the high temperature case.
However, if temperature is further lowered
($T\ll T_K^{(0)}$) for $\veps_d\ll E_F$, charge fluctuations would be quenched
and the Kondo resonance would be pinned at the Fermi level,
independently of $\varphi$ and $\mathcal{T}_b$.
As a consequence, the magnetoasymmetry would tend to vanish. 

\section{Shot noise}\label{sec_noise}
The shot noise is a valuable tool in the characterization of the
transport properties of mesoscopic systems.\cite{bla00}
For systems described with Anderson impurity models like ours,
the electron repulsion term $U$ introduces correlations which
can be investigated through the noise. Then, the problem becomes involved,
although the effect of Kondo correlations in the shot noise have been already
addressed in a number of papers.\cite{her92,yam94,din97,mei02,don02,%
avi03,aon03,lop03a,lop03b,san05b,wu05,gol06,sel06,zha07,zar07}.

Magnetoasymmetries in noise have recently attracted a good deal
of attention due to the (weakly) nonequilibrium relations between the asymmetries
of the current and that of the noise to leading order
in a voltage expansion.\cite{foe08,sai08,san09,uts09} The subject is also of interest because
it poses questions about the validity of fluctuation theorems
out of equilibrium.\cite{foe08} In this section our goal is to calculate
the noise power for our system in the limits of both weak
and strong electron-electron interactions
and check the nonequilibrium fluctuation relations.

The current noise between terminals $\alpha$ and $\beta$
is defined as
\beq\label{eq:noise}
S_{\alpha\beta}(t-t')
= \frac{1}{2} \left\{ \nbraket{[\hat{I}_{\alpha}(t),\hat{I}_{\beta}(t')]_+} - 2\nbraket{\hat{I}_{\alpha}}\nbraket{\hat{I}_{\beta}} \right\} \,,
\edq
where $\hat{I}$ represents a current operator.
The Fourier transformation of the current noise (the noise power) reads
\beq
S(\omega) \equiv \int_{-\infty}^{\infty} dt~ e^{i\omega t} S(t) \,.
\edq
where we have defined the Fourier transform without the prefactor $2$.
In the following, we present results for the zero frequency case
[the shot noise $S \equiv S(0)$].

The noise definition of Eq.~(\ref{eq:noise}) contains correlations between
currents that, quite generally, involve four operators. To treat the resulting
two-body Green's functions, we make use of cluster expansion,\cite{lop03a,san05b,zha07}
\begin{multline}
\nbraket{\hatO_{\mu\sm}^{\dag}\hatO_{\nu\sm}\hatO_{\mu'\sm'}^{\dag}\hatO_{\nu'\sm'}}
\approx \nbraket{\hatO_{\mu\sm}^{\dag}\hatO_{\nu\sm}}\nbraket{\hatO_{\mu'\sm'}^{\dag}\hatO_{\nu'\sm'}}
\\
+ \delta_{\sm\sm'} \nbraket{\hatO_{\mu\sm}^{\dag}\hatO_{\nu'\sm'}}\nbraket{\hatO_{\nu\sm}\hatO_{\mu'\sm'}^{\dag}}
\label{eq:clusterexp}
\end{multline}
which amounts to neglecting two-body connected Green's functions.
As a result, the shot noise is expressed in terms of 
one-body Green's functions. This is a strong assumption that can
lead to deviations from well known relations. Nevertheless,
interactions are included (at the level of the Lacroix's approximation).
The calculation is lengthy and we refer the reader to App.~C.
Here, we consider limit cases only.
\begin{figure}
\centering
\includegraphics[width=0.45\textwidth]{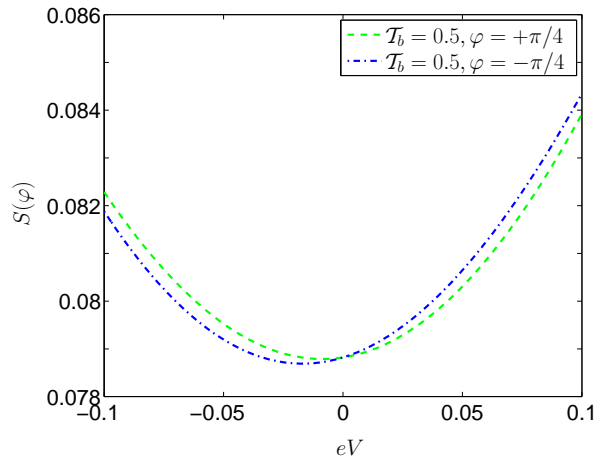} 
\caption{(Color online) Noise ($2e^2/h$) as a function of voltage for
two opposite orientations of $\varphi$.
Parameters are $D = 1$, $\veps_d = -0.05$, $U = \infty$, and $\Gamma_L = \Gamma_R = 0.031$.}
\label{fig:noise}
\end{figure}

For the noninteracting case and at zero temperature we recover the known expression,
\beq
S = \frac{e^2}{h} \sum_\sigma \int_{-eV/2}^{eV/2}
d\veps~ \calt_\sigma(\omega) \left[1-\calt_\sigma(\omega)\right] \,,
\label{eq:shotnoiseT0}
\edq
where the transmission is given by Eq.~\eqref{eq_caltfano}.
As expected, the noise is an even function of $\varphi$ to all orders
in $V$. However, interactions destroy this symmetry already
in the linear regime of the noise response. In Fig.~\ref{fig:noise}
we show the noise as a function of $V$
in the strongly interacting case.
We can observe that the slope of the noise curves
at $V=0$ differ for opposite field orientations.
Another interesting feature is that for some voltages
the nonequilibrium  noise can be {\em reduced} from its equilibrium value.\cite{for09}
\begin{figure}
\centering
\includegraphics[width=0.45\textwidth]{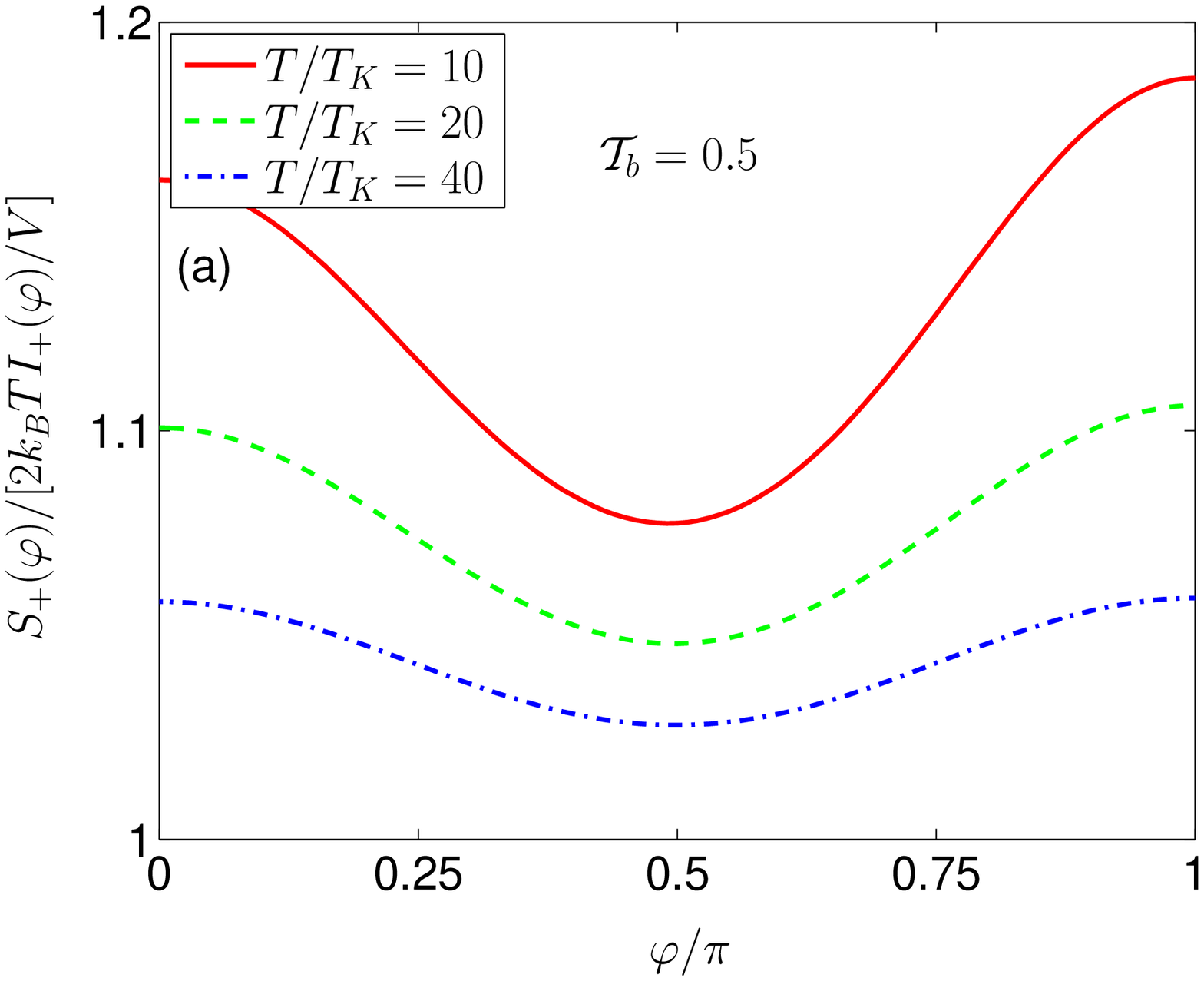} 
\includegraphics[width=0.45\textwidth]{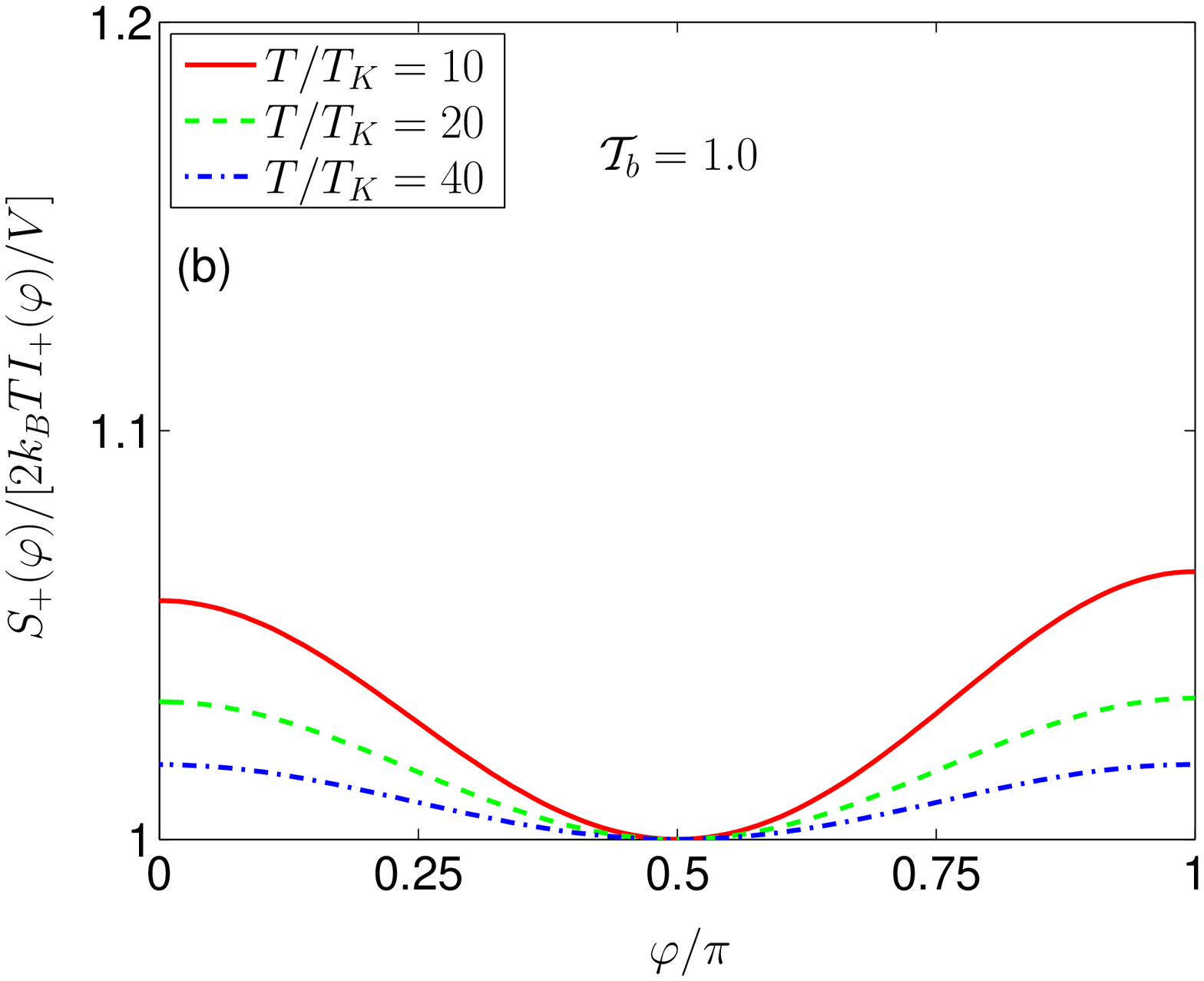}
\caption{(Color online)
Identification of the fluctuations-dissipation theorem as a function of $\varphi$.
Parameters are $D = 1$, $\veps_d = -0.05$, $U = \infty$, and $\Gamma_L = \Gamma_R = 0.031$.
The applied bias is $eV = 0.01T_K^{(0)}$.}
\label{fig:Kubo}
\end{figure}

To gain further insight, we expand the noise in powers of $V$,
\beq
S=S_0+S_1V+\ldots\,.
\edq
$S_0$ is the equilibrium noise describing thermal fluctuations.
Since these fluctuations do not distinguish between $+\varphi$
and $-\varphi$, $S_0$ is an even function of the magnetic field.
An alternate proof of this statement is based on the equilibrium
fluctuation-dissipation theorem,
which relates $S_0$ to the linear conductance $G_0$,
\beq\label{eq:fdt}
S_0=2k_B T G_0\,.
\edq
Since for $G_0$ the Onsager symmetry holds, $S_0$ should be even
for a two-terminal setup.

We now show the numerical results of our model for different
temperatures. We define the symmetric and antisymmetric components
of the noise and the current as before,
\bes
\begin{align}
S_{\pm}&=S(+\varphi) \pm S(-\varphi)\,,\\
I_{\pm}&=I(+\varphi) \pm I(-\varphi)\,.
\end{align}
\eds
We consider symmetric couplings. As result, the even-odd properties
of the transport coefficients allow us to write,
\bes
\begin{align}
\label{eqspm}\frac{S_+}{I_+/V}&=2k_B T\,\\
\label{eqspm2}\frac{S_-}{I_-/V^2}&=2k_B T\,.
\end{align}
\eds
Corrections to these relations are of order $V^2$
and can be neglected for $eV\ll k_B T$.
The first relation is merely a restatement of Eq.~(\ref{eq:fdt}).
The second relation is a nonequilibrium relation
that connect the magnetoasymmetries corresponding to both
the linear-response noise ($S_1$)
and the leading-order nonlinear conductance ($G_1$).
\begin{figure}
\centering
\includegraphics[width=0.45\textwidth]{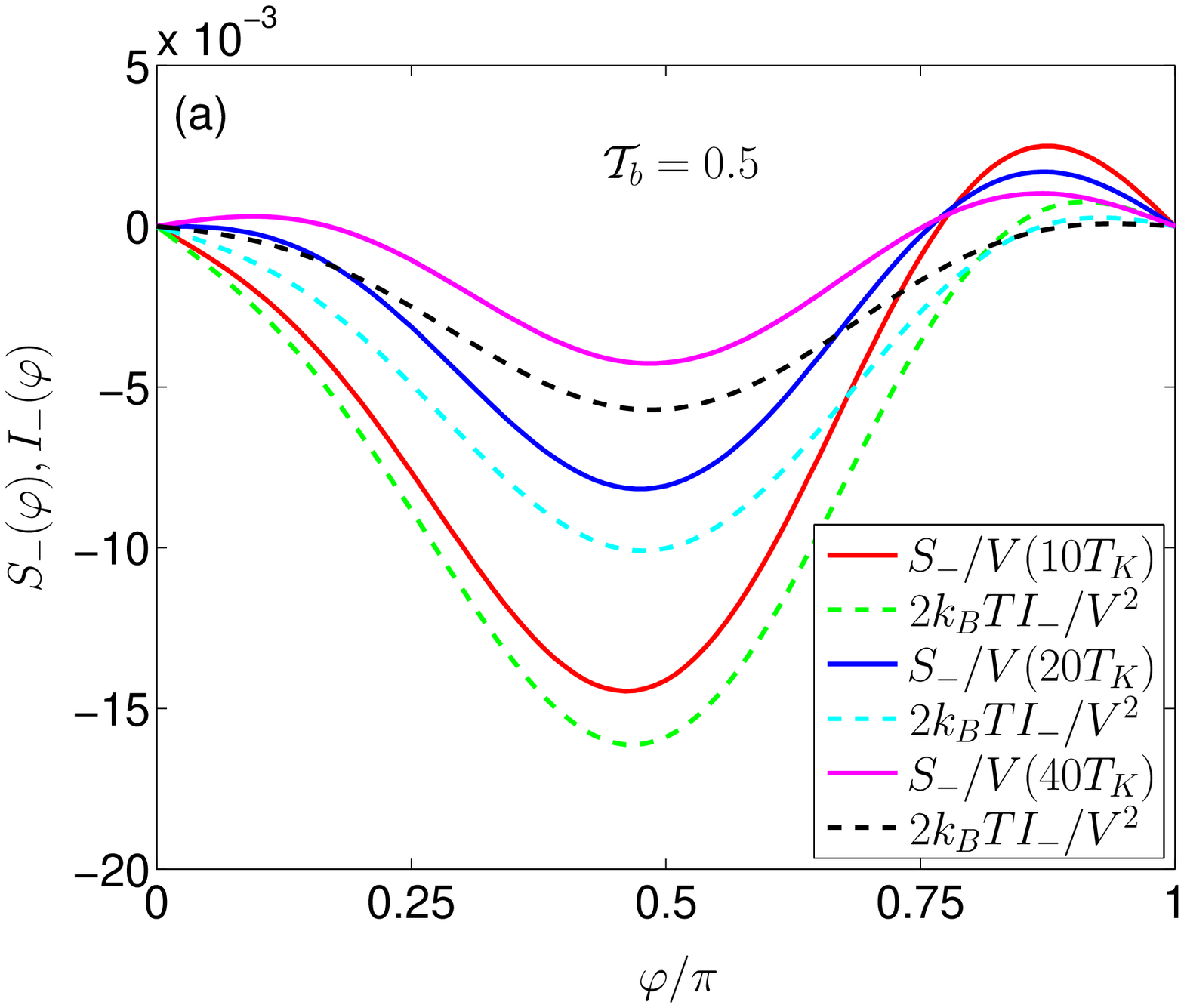}
\includegraphics[width=0.45\textwidth]{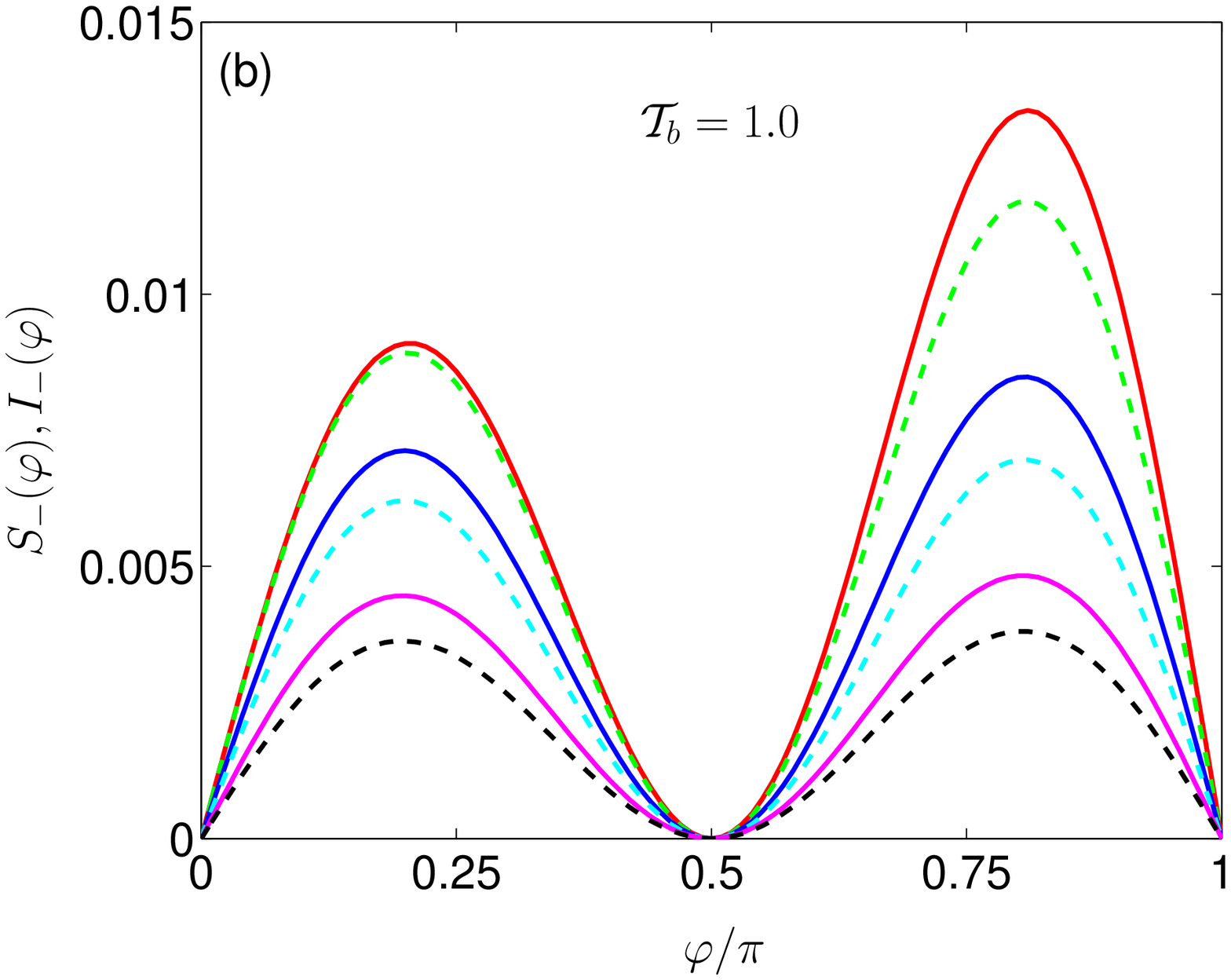}
\caption{(Color online) Antisymmetric parts of noise,
$S_{-}(\varphi)$, and current, $I_{-}(\varphi)$, as a function of $\varphi$.
Parameters are $D = 1$, $\veps_d = -0.05$, $U = \infty$, and $\Gamma_L = \Gamma_R = 0.031$.
The applied bias is $eV = 0.01T_K^{(0)}$.}
\label{fig:smim}
\end{figure}

We take a small voltage ($V = 0.01T_K^{(0)}/e$) and
plot in Fig.~\ref{fig:Kubo} the relation given by Eq.~\eqref{eqspm}
as a function of the flux for two different values of
the background transmission. In bot cases we find that Eq.~\eqref{eqspm}
is approximately fulfilled with small deviations which we attribute
to the assumption of Eq.~(\ref{eq:clusterexp}).
We note that deviations grow for smaller temperatures since
in this case the Kondo correlations become more relevant
and our model for the noise starts to break down.
When the nonresonant channel is fully open
(bottom panel of Fig.~\ref{fig:Kubo}), the deviations are less important
since electrons preferably travel along the upper arm
and consequently feel less the intradot interactions.

The validity of the nonequilibrium fluctuation relation [Eq.~\eqref{eqspm2}]
is analyzed in Fig.~\ref{fig:smim}.
Here we plot separately the two terms of Eq.~\eqref{eqspm2}.
We find a strong ressemblance between $S_1$ and $G_1$ for all
magnetic fields. Deviations also exist as in the calculation of the
symmetric components but they fulfill the same pattern, namely,
they tend to disappear when the background transmission is close
to 1 and the temperature increases well above the Kondo temperature.
Although our results are not a conclusive proof of Eq.~\eqref{eqspm2}
for strongly interacting systems, the errors are small and compatible
with the same deviations found in the equilibrium case (Fig.~\ref{fig:Kubo}).

\section{Conclusions}\label{sec_concl}
We have shown that the current--voltage characteristics
of a two-terminal quantum-dot mesoscopic interferometer
is not an even function of the applied flux in
the nonlinear regime of transport and when
intradot interactions are taken into account.
The interference pattern of the ring at a finite bias
is thus not symmetric under reversal of the magnetic
field. We have carefully investigated the symmetry
properties of the conductance coefficients in a
current--voltage expansion.
Our discussions are based on the properties
of the charge response of the dot when
a finite bias is applied to the system.
When the quantum dot
is in the Coulomb-blockade regime, we find
for most cases that the even coefficients are
symmetric functions of the field while the
odd coefficients does not show any relevant
symmetry in the general case. Only when the dot
is symmetrically coupled to the leads the
odd coefficients are antisymmetric.

We have also calculated the magnetoasymmetry of the
system in the strong coupling regime, when the dot
is described with Kondo correlations. In this case,
the magnetoasymmetry shows an abrupt transition
between positive and negative values when
the voltage crosses the Fermi energy. As a result,
the Kondo resonance dominates the magnetoasymmetry
lineshape when the voltage is of the order of the
Kondo temperature. A further extension of this work
could be focused on the very low temperature
regime using, e.g., slave-boson techniques.

Finally, we have investigated the asymmetry
in the shot noise, finding a correlation between
the noise and the current magnetoasymmetry to leading order
in the applied voltage. This nonequilibrium
fluctuation relation seems to apply in
a wide range of parameters (temperature,
direct channel transmission and applied fluxes).
However, further work is needed to reduce
the deviations which are most probably due
to our approximations. In particular, it would
be interesting to analyze the role of the
third cumulant of the current (or better
the entire full counting statistics) under
field reversal.

\section*{Acknowledgements}
This work was supported by the Spanish MICINN Grant No.\ FIS2008-00781
and the Conselleria d'Innovaci\'o, Interior i Justicia
(Govern de les Illes Balears).

\appendix

\section{Lesser Green's Function of the Dot}

In general, the lesser Green's function for an interacting dot cannot be obtained from the equation of motion technique without introducing additional assumptions.
Following Ng's heuristic approach, here we employ an ansatz for interacting lesser and greater Green's functions \cite{Ng96}.
Using the lesser and greater self-energies for a noninteracting dot
\bes
\begin{align}
\Sigma_{\sm}^{<(0)}(\omega) &= +i\sum_{\alpha} \bar{\Gamma}_{\alpha}f_{\alpha}(\omega) \,,
\\
\Sigma_{\sm}^{>(0)}(\omega) &= -i\sum_{\alpha} \bar{\Gamma}_{\alpha}\left[1 - f_{\alpha}(\omega)\right] \,,
\end{align}
\eds
with
\bes
\begin{align}
\bar{\Gamma}_L &= \frac{1}{(1+\xi)^2}\left(\Gamma_L + \xi\Gamma_R\right) + \frac{\wG}{2}\sqrt{\alpha\calt_b}\sin(\varphi) \,,
\\
\bar{\Gamma}_R &= \frac{1}{(1+\xi)^2}\left(\Gamma_R + \xi\Gamma_L\right) - \frac{\wG}{2}\sqrt{\alpha\calt_b}\sin(\varphi) \,.
\end{align}
\eds
we assume that lesser and greater Green's functions for an interacting dot can be written in the form
\bes
\begin{align}
\Sigma_{\sm}^<(\omega) &= +i\sum_{\alpha} \bar{\Gamma}_{\alpha} f_{\alpha}R_{\sm}(\omega) \,, \\
\Sigma_{\sm}^>(\omega) &= -i\sum_{\alpha} \bar{\Gamma}_{\alpha} \left[1-f_{\alpha}\right] R_{\sm}(\omega) \,.
\end{align}
\eds
The explicit form of $R_{\sm}(\omega)$ can be obtained from the relation
\beq
\Sigma_{\sm}^{>}(\omega) - \Sigma_{\sm}^{<}(\omega) = \Sigma_{\sm}^{r}(\omega) - \Sigma_{\sm}^{a}(\omega) \,.
\edq
Employing the identity,
\beq
\calg_{d\sm,d\sm}^{>}(\omega) - \calg_{d\sm,d\sm}^{<}(\omega) = \calg_{d\sm,d\sm}^{r}(\omega) - \calg_{d\sm,d\sm}^{a}(\omega) \,,
\edq
it finally yields
\beq
\calg_{d\sm,d\sm}^{<}(\omega) = -f_{peq}(\omega)\left[\calg_{d\sm,d\sm}^r(\omega) - \calg_{d\sm,d\sm}^a(\omega) \right] \,,
\edq
with
\beq
f_{peq}(\omega) = \frac{\sum_{\alpha} \bar{\Gamma}_{\alpha}f_{\alpha}(\omega)}{\sum_{\alpha} \bar{\Gamma}_{\alpha}} \,.
\edq

\section{Evaluation of expectation values} \label{app:expeval}

In deriving Eq.~\eqref{eq:greensc}, we have to evaluate the expectation values 
$\nbraket{d_{\bsm}^{\dag}c_{\alpha k\bsm}}$ and $\sum_{\beta,r} V_{\beta} \nbraket{c_{\beta r\bsm}^{\dag}c_{\alpha k\bsm}}$.
First, let us concentrate on $\nbraket{d_{\bsm}^{\dag}c_{\alpha k\bsm}}$.
In equilibirum, the quantities can be calculated by using the fluctuation-dissipation theorem,
\beq
\nbraket{d_{\bsm}^{\dag}c_{\alpha k\bsm}}
= - \frac{1}{\pi} \int d\omega~ f_{eq}(\omega) \Im\left[\dbraket{c_{\alpha k\bsm},d_{\bsm}^{\dag}}_{\omega}^r\right] \,.
\edq
However, the fluctuation-dissipation theorem cannot be employed in nonequilibrium.
Then, in general, 
\beq
\nbraket{d_{\bsm}^{\dag}c_{\alpha k\bsm}}
= \frac{1}{2\pi i}\int d\omega~ \dbraket{c_{\alpha k\bsm},d_{\bsm}^{\dag}}_{\omega}^< \,.
\edq
Note that $\dbraket{c_{\alpha k\bsm},d_{\bsm}^{\dag}}_{\omega}^<$ cannot be obtained exactly becuase of the dot lesser Green's function.
Here, we thus assume the following pseudoequilibrium form
\beq
\dbraket{c_{\alpha k\bsm},d_{\bsm}^{\dag}}_{\omega}^< = -f_{peq}(\omega) 
\left(\dbraket{c_{\alpha k\bsm},d_{\bsm}^{\dag}}_{\omega}^r - \dbraket{c_{\alpha k\bsm},d_{\bsm}^{\dag}}_{\omega}^a\right) \,.
\edq
Then, we have
\begin{multline}
\nbraket{d_{\bsm}^{\dag}c_{\alpha k\bsm}}
= \\
-\frac{1}{2\pi i}\int d\omega~ f_{peq}(\omega) 
\left(\dbraket{c_{\alpha k\bsm},d_{\bsm}^{\dag}}_{\omega}^r(\varphi) - \dbraket{c_{\alpha k\bsm},d_{\bsm}^{\dag}}_{\omega}^a(\varphi)\right) \,,
\end{multline}
where
\bes
\begin{align}
\dbraket{c_{Lk\bsm},d_{\bsm}^{\dag}}_{\omega}^r 
&= \frac{g_{Lp\bsm}^r}{1+\xi} \left( V_L - i\sqrt{\xi}e^{-i\varphi}V_R \right)
\dbraket{ d_{\bsm},d_{\bsm}^{\dag}}_{\omega}^r \,, \\
\dbraket{c_{Rk\bsm},d_{\bsm}^{\dag}}_{\omega}^r
&= \frac{g_{Rq\bsm}^r}{1+\xi} \left( V_R - i\sqrt{\xi}e^{+i\varphi}V_L \right)
\dbraket{ d_{\bsm},d_{\bsm}^{\dag}}_{\omega}^r \,,
\end{align}
\eds
and
\bes
\begin{align}
\dbraket{c_{Lp\bsm},d_{\bsm}^{\dag}}_{\omega}^a &= \frac{g_{Lp\bsm}^a}{1+\xi} \left(V_L + i\sqrt{\xi}e^{-i\varphi}V_R\right) \dbraket{d_{\bsm},d_{\bsm}^{\dag}}_{\omega}^a \,,
\\
\dbraket{c_{Rq\bsm},d_{\bsm}^{\dag}}_{\omega}^a &= \frac{g_{Rq\bsm}^a}{1+\xi} \left(V_R + i\sqrt{\xi}e^{+i\varphi}V_L\right) \dbraket{d_{\bsm},d_{\bsm}^{\dag}}_{\omega}^a \,,
\end{align}
\eds
with $g_{\alpha k\bsm}^{r,a} = 1/(\omega \pm i0^+ - \veps_{\alpha k\bsm})$.
Due to the approximations, however, the retarded Green's function is not
the complex conjugate of the advanced one so that this equation gives an unphysical result.
To resolve this difficulty, we thus replace the quantity in the parenthesis by the time-reversal pair. That is,
\begin{multline}
\nbraket{d_{\bsm}^{\dag}c_{\alpha k\bsm}}
= -\frac{1}{2\pi i}\int d\omega~ f_{peq}(\omega) 
\\
\times \left(\dbraket{c_{\alpha k\bsm},d_{\bsm}^{\dag}}_{\omega}^r(-\varphi) - \dbraket{c_{\alpha k\bsm},d_{\bsm}^{\dag}}_{\omega}^a(\varphi)\right) \,.
\label{eq:pseudocd}
\end{multline}
This yields a real expectation value as should be.
Using Eq.~\eqref{eq:pseudocd}, we thus have
\bes
\begin{multline}
\sum_k \frac{\nbraket{d_{\bsm}^{\dag}c_{Lk\bsm}}}{\omega + i0^+ - \veps_{Lk\bsm}} 
= -\frac{\rho_0}{1+\xi} \int d\omega'~ \frac{f_{peq}(\omega')}{\omega' - \omega - i0^+} 
\\
\times \left(V_L + i\sqrt{\xi}e^{-i\varphi}V_R\right)
\left[\dbraket{d_{\bsm},d_{\bsm}^{\dag}}_{\omega'}^r\right]^{\ast} \,,
\end{multline}
\begin{multline}
\sum_k \frac{\nbraket{d_{\bsm}^{\dag}c_{Rk\bsm}}}{\omega + i0^+ - \veps_{Rk\bsm}} 
= -\frac{\rho_0}{1+\xi} \int d\omega'~ \frac{f_{peq}(\omega')}{\omega' - \omega - i0^+} 
\\
\times \left(V_R + i\sqrt{\xi}e^{+i\varphi}V_L\right)
\left[\dbraket{d_{\bsm},d_{\bsm}^{\dag}}_{\omega'}^r\right]^{\ast} \,.
\end{multline}
\label{eq:cdden}
\eds
In the same way:
\bes
\begin{multline}
\sum_{\beta,r,k} \frac{\nbraket{c_{\beta r\bsm}^{\dag}c_{Lk\bsm}}}{\omega + i0^+ - \veps_{Lk\bsm}} 
= -\frac{\rho_0}{1+\xi} \int d\omega'~ \frac{f_{peq}(\omega')}{\omega' - \omega - i0^+} 
\\
\times \left(V_L + i\sqrt{\xi}e^{-i\varphi}V_R\right)
\left\{1 + \left[\Sigma_0(\omega')\dbraket{d_{\bsm},d_{\bsm}^{\dag}}_{\omega'}^r\right]^{\ast}\right\} \,,
\end{multline}
\begin{multline}
\sum_{\beta,r,k} \frac{\nbraket{c_{\beta r\bsm}^{\dag}c_{Rk\bsm}}}{\omega + i0^+ - \veps_{Rk\bsm}} 
= -\frac{\rho_0}{1+\xi} \int d\omega'~ \frac{f_{peq}(\omega')}{\omega' - \omega - i0^+} 
\\
\times \left(V_R + i\sqrt{\xi}e^{+i\varphi}V_L\right)
\left\{1 + \left[\Sigma_0(\omega')\dbraket{d_{\bsm},d_{\bsm}^{\dag}}_{\omega'}^r\right]^{\ast}\right\} \,.
\end{multline}
\label{eq:ccden}
\eds

\section{Expression of the shot noise}

The current noise is defined as
\beq
S_{\alpha\beta}(t-t')
= \frac{1}{2} \left\{ \nbraket{[\hat{I}_{\alpha}(t),\hat{I}_{\beta}(t')]_+} - 2\nbraket{\hat{I}_{\alpha}}\nbraket{\hat{I}_{\beta}} \right\}\,.
\edq
The current operator can be calculated from the time evolution of the occupation number operator
\begin{multline}
\hat{I}_R = \frac{ie}{\hbar} \left\{ 
\sum_{p\in L, q\in R,\sigma} \left[ We^{i\varphi} c_{Rq\sigma}^{\dag}c_{Lp\sigma} - We^{-i\varphi}c_{Lp\sigma}^{\dag}c_{Rq\sigma} \right] 
\right.
\\
\left.
+  \sum_{q\in R,\sigma} \left[ V_R c_{Rq\sigma}^{\dag}d_{\sigma} - V_R^{\ast} d_{\sigma}^{\dag}c_{Rq\sigma} \right] \right\}\,.
\label{eq:cop}
\end{multline}
The current can be then obtained by taking the average of $\hat{I}_R$.
Using the Keldysh Green's functions, the current can be expressed as
\begin{multline}
\nbraket{\hat{I}_R(t)}
= \frac{e}{h} \left[ \sum_{p\in L, q\in R,\sigma}  
\left( We^{i\varphi} \calg_{Lp\sigma,Rq\sigma}^{<}(t,t) 
\right.\right.
\\
\left.\left.
- We^{-i\varphi} \calg_{Rq\sigma,Lp\sigma}^<(t,t) \right)
\right. 
\\
\left.
+  \sum_{q\in R,\sigma} \left( V_R \calg_{d\sigma,Rq\sigma}^{<}(t,t) - V_R^{\ast}\calg_{Rq\sigma,d\sigma}^<(t,t) \right) \right]\,.
\label{eq:cmop}
\end{multline}

The Fourier transformation of the current noise is
\beq
S(\omega) \equiv \int_{-\infty}^{\infty} dt~ e^{i\omega t} S(t)\,.
\edq
The zero-frequency noise power $S \equiv S(0)$ is referred to as shot noise.
In the derivation, we make use of cluster expansion to treat two-body Green's functions:\cite{lop03a,san05b,zha07}
\begin{multline}
\nbraket{\hatO_{\mu\sm}^{\dag}\hatO_{\nu\sm}\hatO_{\mu'\sm'}^{\dag}\hatO_{\nu'\sm'}}
\approx \nbraket{\hatO_{\mu\sm}^{\dag}\hatO_{\nu\sm}}\nbraket{\hatO_{\mu'\sm'}^{\dag}\hatO_{\nu'\sm'}}
\\
+ \delta_{\sm\sm'} \nbraket{\hatO_{\mu\sm}^{\dag}\hatO_{\nu'\sm'}}\nbraket{\hatO_{\nu\sm}\hatO_{\mu'\sm'}^{\dag}}
\end{multline}
In the frequency domain, using Eqs.~\eqref{eq:cop} and \eqref{eq:cmop} and cluster expansion the shot noise is given by
\begin{widetext}
\begin{multline}
S = \frac{1}{2\pi} \frac{2(ie)^2}{\hbar}  \int d\omega~ \left\{ 
W^2 \left[e^{+2i\varphi} \calg_{Lp'\sm,Rq\sm}^<(\omega) \calg_{Lp\sm,Rq'\sm}^>(\omega) 
+ e^{-2i\varphi} \calg_{Rq'\sm,Lp\sm}^<(\omega) \calg_{Rq\sm,Lp'\sm}^>(\omega) \right] \right. \\
- W^2 \left[\calg_{Rq'\sm,Rq\sm}^<(\omega) \calg_{Lp\sm,Lp'\sm}^>(\omega) + \calg_{Lp'\sm,Lp\sm}^<(\omega) \calg_{Rq\sm,Rq'\sm}^>(\omega) \right] \\
+ W \left[ V_R e^{+i\varphi} \calg_{d\sm,Rq\sm}^<(\omega)\calg_{Lp\sm,Rq'\sm}^>(\omega) 
+ V_R^{\ast} e^{-i\varphi} \calg_{Rq'\sm,d\sm}^<(\omega)\calg_{Rq\sm,Lp'\sm}^>(\omega) \right] \\
- W \left[ V_R e^{-i\varphi} \calg_{d\sm,Lp\sm}^<(\omega)\calg_{Rq\sm,Rq'\sm}^>(\omega) 
+ V_R^{\ast} e^{+i\varphi} \calg_{Lp'\sm,d\sm}^<(\omega)\calg_{Rq\sm,Rq'\sm}^>(\omega) \right] \\
- W \left[ V_R^{\ast} e^{+i\varphi} \calg_{Rq'\sm,Rq\sm}^<(\omega)\calg_{Lp\sm,d\sm}^>(\omega) 
+ V_R e^{-i\varphi} \calg_{Rq'\sm,Rq\sm}^<(\omega)\calg_{d\sm,Lp'\sm}^>(\omega) \right] \\
+ W \left[ V_R^{\ast} e^{-i\varphi} \calg_{Rq'\sm,Lp\sm}^<(\omega)\calg_{Rq\sm,d\sm}^>(\omega) 
+ V_R e^{+i\varphi} \calg_{Lp'\sm,Rq\sm}^<(\omega)\calg_{d\sm,Rq'\sm}^>(\omega) \right] \\
+ \left[ V_R V_R \calg_{d\sm,Rq\sm}^<(\omega)\calg_{d\sm,Rq'\sm}^>(\omega) 
+ V_R^{\ast} V_R^{\ast} \calg_{Rq'\sm,d\sm}^<(\omega)\calg_{Rq\sm,d\sm}^>(\omega) \right] \\
\left. - V_R V_R^{\ast} \left[ \calg_{Rq'\sm,Rq\sm}^<(\omega)\calg_{d\sm,d\sm}^>(\omega) 
+ \calg_{d\sm,d\sm}^<(\omega)\calg_{Rq\sm,Rq'\sm}^>(\omega) \right] \right\} \,,
\label{eq:shotnoiseexp}
\end{multline}
\end{widetext}
where summations over momentum indices are assumed.
Here, the Keldysh Green's functions which appear on the r.h.s. of Eq.~\eqref{eq:shotnoiseexp} can be expressed in terms of the dot Green's functions, 
$\calg_{d\sm,d\sm}^<(\omega)$ and $\calg_{d\sm,d\sm}^{r,a}(\omega)$.


\begin{thebibliography}{90}
\bibitem{zeeman}
We have neglected the Zeeman interaction due to the
applied magnetic field since in experimentally available
mesoscopic interferometers the applied fields are usually
quite small.\cite{let06}
\end{thebibliography}
\end{document}